\documentclass[twocolumn,final]{autart}    
\usepackage{ifpdf}
\usepackage{amsmath,amssymb,amsfonts}
\usepackage{url}
\usepackage{graphics} 
\usepackage{graphicx}
\usepackage{epsfig}
\usepackage{stfloats}
\usepackage{mathtools}
\usepackage{color,soul} 
\usepackage{enumitem}
\usepackage[authoryear]{natbib}
\usepackage{xcolor}
\usepackage[colorlinks=true,linkcolor=blue, citecolor=blue, urlcolor=blue]{hyperref}
\allowdisplaybreaks
\usepackage{easyReview} 
\usepackage{comment}

\usepackage{placeins}

\newtheorem{assum}{Assumption}

\newcommand{\SO}{\operatorname{SO}}
\renewcommand{\so}{\mathfrak{so}}
\newcommand{\SE}{\operatorname{SE}}
\newcommand{\se}{\mathfrak{se}}

\DeclareMathOperator{\col}{col}
\DeclareMathOperator{\tr}{Tr}
\DeclareMathOperator{\dom}{dom}  
\DeclareMathOperator{\Ad}{Ad}  
\DeclareMathOperator{\ad}{ad}

\DeclareMathOperator{\blkdiag}{blkdiag}
\DeclareMathOperator{\diag}{diag}
\DeclareMathOperator{\nulls}{Null}
\DeclareMathOperator{\rank}{rank}
\DeclareMathOperator{\kg}{kg}
\DeclareMathOperator{\m}{m}
\newcommand{\T}{^{\top}} 
\newcommand{\ie}{\textit{i.e.}}

\allowdisplaybreaks
\AtBeginDocument{%
	\setlength{\abovedisplayskip}{9pt}
	\setlength{\belowdisplayskip}{9pt}
	\setlength{\abovedisplayshortskip}{9pt}
	\setlength{\belowdisplayshortskip}{9pt}
}

\begin{document}
	
	\begin{frontmatter}
		
		\title{Distributed Global Pose Synchronization for Heterogeneous Multiple Rigid Body Systems on SE(3)} 
		
		\thanks[footnoteinfo]{This work was supported in part by the National Natural Science Foundation of China under Grant 62403205, 62425602 and 62595802, and in part by the National Sciences and Engineering Research Council of Canada, under Grant NSERC-DG RGPIN-2026-06037. A preliminary version of this work has been presented at the IFAC World Congress 2026 \citep{lin2026IFAC}.}
		
		\author[First]{Fengyu Lin}\ead{linfy2024@hust.edu.cn},    
		\author[First]{Miaomiao Wang}\ead{mmwang@hust.edu.cn},               
		\author[First]{Housheng Su}\ead{shs@hust.edu.cn},
		\author[First]{Zhigang Zeng}\ead{zgzeng@hust.edu.cn},
		\author[Second]{Abdelhamid Tayebi}\ead{atayebi@lakeheadu.ca}
		
		\address[First]{School of Artificial Intelligence and Automation, Huazhong University of Science and Technology, Wuhan 430074, China}             
		\address[Second]{Department of Electrical Engineering, Lakehead University, Canada, and School of Mathematics, Shandong University, China}        


		\begin{abstract}    
			This paper addresses the problem of global pose synchronization for heterogeneous multiple rigid-body systems evolving on the matrix Lie group $\SE(3)$. We first construct a new potential function on $\SE(3) \times \mathbb{R}$ with a generalized non-diagonal weighting matrix, together with a set of switching variables governed by continuous-discrete hybrid dynamics. Based on this new potential function and these switching variables, we propose a distributed hybrid feedback control scheme that uses relative pose on $\SE(3)$ and group velocity measurements, endowed with global asymptotic stability guarantees. We then further propose a distributed velocity-free hybrid feedback control scheme that uses relative and absolute pose measurements, without requiring group velocity measurements, while preserving global asymptotic stability. Numerical simulations are presented to demonstrate the performance of the proposed distributed hybrid control schemes.
		\end{abstract}
		
	\end{frontmatter}
	\section{Introduction}
	The pose (attitude and position) synchronization control problem of multiple rigid body systems (MRBSs) has been widely investigated over the past few decades due to its significant importance in many practical applications, such as robotics \citep{bullo2009distributed}, satellites \citep{subbarao2008nonlinear}, unmanned aerial vehicles (UAVs) \citep{rashad2020fully} and spacecraft \citep{zou2012attitude}.
	Several approaches have been proposed to address the pose synchronization problem of rigid body systems. A commonly adopted strategy is the decoupled control framework on $\SO(3) \times \mathbb{R}^3$ resulting from the semidirect product of the rotation group $SO(3)$ and the translation group $\mathbb{R}^3$, which is a smooth structure diffeomorphic to the Lie group $SE(3)$ \citep{gallier2011geometric}. The key idea is to decouple the rigid-body dynamics into independent (or cascaded) rotational and translational subsystems and then design feedback control laws for attitude and position separately (see, for instance, \citet{nair2007stable}). However, these decoupled approaches fundamentally neglect the intrinsic geometric coupling between rotational and translational dynamics inherent to $\SE(3)$, which may lead to weak stability guarantees. Alternatively, approaches based on dual quaternions, which allow for representing the pose in a compact form without requiring decoupling, have been proposed in \cite{wang2012dual,malladi2020rigid}. However, the dual quaternions representation of $SE(3)$ is a non-unique representation (double cover of the space of rotations) and may result in the so-called unwinding phenomenon in practice \citep{mayhew2012quaternion}.
	Few works consider the simultaneous coordination of the attitude and position dynamics on the Lie group $\SE(3)$, such as \cite{thunberg2016consensus,hatanaka2012passivity}. The method proposed in \cite{thunberg2016consensus}, leverages the angle-axis representation of the rotation to design the control law. Moreover, in \cite{hatanaka2012passivity}, a passivity-based pose/attitude synchronization approach on $\SE(3)$/$\SO(3)$ have been proposed. However, the synchronization control schemes developed in \cite{thunberg2016consensus,hatanaka2012passivity} fail to guarantee global asymptotic stability (GAS) due to the topological properties of the motion space containing $\SO(3)$.  In fact, almost global asymptotic stability (AGAS)\footnote{AGAS refers to the fact that the equilibrium point is stable and attractive from all initial conditions except a set of zero Lebesgue measure.} is the strongest stability result that one can achieve, with time-invariant continuous feedback, in spaces involving $\SO(3)$.  Since $\SE(3)$ is diffeomorphic to $\SO(3) \times \mathbb{R}^3$, it inherits the topology of $\SO(3)$, which precludes global asymptotic stabilization by continuous feedback.
	
	To overcome the above-mentioned topological obstruction to GAS on $\SO(3)$ and $\SE(3)$, a hybrid feedback framework \citep{goebel2012hybrid}, relying on a “synergistic" family of potential functions on $\SO(3)$, has been proposed in \cite{mayhew2013synergistic} for attitude control, leading to GAS guarantees. This hybrid feedback framework has been extended to solve the global attitude tracking and estimation problems on $SO(3)$, for instance, \cite{mayhew2013synergistic,lee2015global,berkane2017construction}. 
	However, most of these hybrid approaches rely on the compactness of the group $\SO(3)$, which makes them difficult to be extended to the global control problem on non-compact Lie groups such as $\SE(3)$. Recently, a new hybrid feedback strategy, relying on a switching variable with hybrid dynamics and a suitable potential function on $\SO(3)\times \mathbb{R}$ (and $\SE(3)\times \mathbb{R}$), has been proposed in \cite{wang2022hybrid,wang2024nonlinear} for global tracking control on matrix Lie groups $\SO(3)$ (and $\SE(3)$). Inspired by this work, the authors in \cite{boughellaba2026global} proposed a distributed attitude synchronization control scheme on $\SO(3)$ with GAS guarantees. However, the design of pose synchronization schemes directly on $\SE(3)$ with GAS guarantees remains a challenging problem. 
	
	On the other hand, measuring the group velocities of all rigid bodies in large-scale MRBSs can be costly. This motivates the design of feedback controllers without angular velocity measurements, which can reduce sensing requirements and serve as backup strategies in the case of gyroscope degradation or failure. For example, an observer-based approach was developed in \cite{zou2012attitude}, to achieve distributed control without requiring absolute or relative angular velocity measurements. A quaternion-based velocity-free attitude synchronization scheme was proposed in \cite{abdessameud2009attitude} relying on the auxiliary system's approach introduced in \cite{Tayebi2008}. This idea has been leveraged in \cite{boughellaba2026global} for the problem of attitude synchronization on $\SO(3)$ with GAS guarantees. However, the design of pose synchronization schemes on $\SE(3)$ without relying on group velocity (both angular velocity and linear velocity) measurements, while providing GAS guarantees, remains a challenging problem that, to the best of our knowledge, has not yet been addressed in the literature.
	
	\textit{Contributions:} In this work, we investigate the global pose synchronization of a group of rigid body systems on Lie group $SE(3)$ under connected and acyclic undirected graphs. A distributed hybrid feedback scheme for pose synchronization of MRBSs on $SE(3)$, endowed with GAS guarantees, is proposed. This distributed hybrid feedback scheme is then extended to deal with the global pose synchronization problem on $\SE(3)$ without group velocity measurements.
	The main contributions of this work are summarized as follows: 
	\begin{enumerate}
		\item We propose a new potential function on $(\SE(3) \times \mathbb{R})^{M}$, involving a set of switching variables with hybrid dynamics, to facilitate the design of the proposed hybrid pose synchronization schemes. 
		Unlike the potential function constructed in \cite{wang2022hybrid} requiring a block diagonal weighting matrix, our proposed potential function is quite general and allows for a non-block-diagonal weighting matrix. This can accommodate weighting matrices that arise naturally in real-world applications (such as landmark positions), offering greater flexibility in the design of hybrid feedback on $\SE(3)$.
		
		\item We propose a new distributed hybrid feedback control scheme directly on $\SE(3)$ for global pose synchronization under connected and acyclic undirected graphs, while guaranteeing GAS. Compared to the existing works, such as \cite{thunberg2016consensus,hatanaka2012passivity}, which provide convergence guarantees subject to restrictions on the initial conditions, the distributed hybrid feedback scheme proposed in this work leads to stronger stability results. 
		\item We propose a velocity-free distributed hybrid feedback for pose synchronization of MRBSs on $\SE(3)$, with GAS guarantees. The proposed controller requires neither angular nor linear velocity measurements, thereby reducing sensing requirements and dependence on dedicated velocity sensors. This feature is particularly important for lightweight, resource-constrained rigid body systems, such as UAVs, where payload capacity and power consumption are critical design considerations.
	\end{enumerate}
	
	The remainder of this paper is organized as follows: Section \ref{Preliminaries} provides the preliminaries used throughout this paper. Section \ref{Problem Statement} formulates the pose synchronization problem on $\SE(3)$. 
	Section \ref{Potential function design} introduces a generic potential function on $(\SE(3)\times \mathbb{R})^M$.  Section \ref{Hybrid Pose Synchronization Scheme} is devoted to designing a distributed hybrid control scheme for global pose synchronization. Section \ref{velocityfree} extends the proposed hybrid control scheme to the problem of velocity-free global pose synchronization. Finally, Section \ref{simulation} presents some simulation results.
	
	\section{Preliminaries}\label{Preliminaries}
	\subsection{Notations}
	The sets of real, positive real, nonnegative real and natural numbers are denoted by $\mathbb{R}, \mathbb{R}_{> 0},  \mathbb{R}_{\geq 0}$, and $\mathbb{N}$, respectively. 
	We denote by $\mathbb{R}^n$ the $n$-dimensional Euclidean space, and denote by $\mathbb{S}^n:= \{x \in \mathbb{R}^{n+1}|x^{\top} x=1\}$ the set of unit vectors in $\mathbb{R}^{n+1}$. For matrices $A, B \in \mathbb{R}^{m\times n}$, the Euclidean inner product is defined as $\langle\langle A, B \rangle\rangle=\tr(A^{\top} B)$. 
	The Euclidean norm of a vector $x\in \mathbb{R}^n$ is defined as $\|x\|=\sqrt{x^{\top} x}$. The matrix $I_n\in \mathbb{R}^{n\times n}$ denotes the identity matrix. Denote the block diagonal matrix by $\blkdiag(\cdot)$. For a given matrix $A \in \mathbb{R}^{n\times n}$, we define the set of all unit-eigenvectors and the $i$-th pair of eigenvalue and eigenvector of $A$ by $\mathcal{E}_v(A)$ and $(\lambda_i^A, v_i^A)$, respectively. Let $\mathbf{0}$ denote a vector or matrix with compatible dimensions. 
	Let $\mathcal{D}$ be a smooth manifold embedded in $\mathbb{R}^n$ with $T_x\mathcal{D}$ being its tangent space at point $x \in \mathcal{D}$. A continuously differentiable real-valued function $f: \mathcal{D} \rightarrow \mathbb{R}_{\geq 0}$ is called a potential function on $\mathcal{D}$ with respect to a closed set $\mathcal{A}$ if $f(x) >0$ for all $x\notin \mathcal{A}$ and $f(x)=0$ for all $x \in \mathcal{A}$. The gradient of $f$ at point $x\in \mathcal{D}$, denoted by $\nabla_xf(x) \in T_x \mathcal{D}$, is uniquely defined by $\dot{f}(x)=\langle \nabla_{x}f(x), \xi\rangle_x$ for all $\xi \in T_x\mathcal{D}$, where $\langle , \rangle_x: T_x\mathcal{D} \times T_x \mathcal{D} \rightarrow \mathbb{R}$ denotes the Riemannian metric on $\mathcal{D}$.  The point $x\in \mathcal{D}$ is called a critical point of $f$ if $\nabla_xf(x)=0$. 
	The cross product between two vectors $y, z\in \mathbb{R}^3$, can be written as the matrix multiplication $y \times z=y^{\times}z$ with the skew-symmetric matrix given by $y^{\times}=[\col(0, y_3, -y_2)~ \col(-y_3, 0, y_1)~ \col(y_2,-y_1,0)]$ and $\col$ denoting the column vector operator.
	Given a matrix $A = [a_{ij}]_{1\leq i,j \leq 3}\in \mathbb{R}^{3\times 3}$, 
	define the map $\psi(A):= \frac{1}{2}[a_{32}-a_{23}, a_{13}-a_{31}, a_{21}-a_{12}]^{\top}$, such that $\langle \langle A,y^{\times} \rangle \rangle=2y^{\top} \psi(A)$ for all $A\in \mathbb{R}^{3\times 3}, y \in \mathbb{R}^3$.

	\subsection{Special Euclidean Group $\SE(3)$}
	The attitude of a rigid body is represented by a rotation matrix $R$ that belongs to the 3-dimensional Special Orthogonal group given by $ \SO(3):=\{R\in \mathbb{R}^{3\times 3}: R^{\top} R=R R^{\top}=I_3, \det(R)=1\}$. 
	The Lie algebra of $\SO(3)$ is denoted by $\so(3):=\{\Omega \in \mathbb{R}^{3 \times 3}:~ \Omega^{T}=-\Omega\}$. For any $R \in \SO(3) $, we define $|R|_I^2:=\tr(I_3-R)/4 \in [0,1]$ as the normalized Euclidean distance with respect to $I_3$.
	Let the map $\mathcal{R}_a: \mathbb{R}\times \mathbb{S}^2 \rightarrow \SO(3)$ denote the well-known angle-axis parameterization of the attitude, given by 
	\begin{align}\label{Rodrigues}
		\mathcal{R}_a(\theta,u):= I_3+\sin(\theta)u^{\times}+(1-\cos(\theta))(u^{\times})^2
	\end{align}
	with $u \in \mathbb{S}^2$ denoting the rotation axis and $\theta \in \mathbb{R}$ denoting the angle of the rotation about the axis. Let us introduce the following maps $\mathcal{T}:\SO(3)\times \mathbb{R}^3 \rightarrow \mathbb{R}^{4\times 4} $ and $(\cdot)^{\wedge}: \mathbb{R}^6 \rightarrow \mathbb{R}^{4\times 4}$ as 
	$$\mathcal{T}(R,p):=\begin{bmatrix}
		R&p\\
		\mathbf{0}&1
	\end{bmatrix}, 
	\xi^{\wedge}=\begin{bmatrix}
		\omega^{\times}&v\\
		\mathbf{0}&0
	\end{bmatrix} $$
	with $R\in \SO(3), p\in \mathbb{R}^3, \xi=[\omega^{\top},v^{\top}]^{\top}$ and $\omega, v\in \mathbb{R}^3$. The 3-dimensional Special Euclidean group is defined by 
	$$\SE(3):=\left\{X=\mathcal{T}(R,p) \in \mathbb{R}^{4\times 4}: R\in \SO(3), p\in \mathbb{R}^3 \right\}
	$$ 
	and the Lie algebra of $\SE(3)$, denoted by $\se(3)$, is defined by 
	$$\se(3):=\left\{U= \xi^{\wedge} \in \mathbb{R}^{4\times 4}: \xi=[\omega^{\top},v^{\top}]^{\top}, \omega,v \in \mathbb{R}^3\right\}.$$
	For any $X \in \SE(3)$, the tangent space is given by $T_X\SE(3)=\{XU \in \mathbb{R}^{4 \times 4}:X \in \SE(3), U \in \se(3)\}$. The inner product in the tangent space $T_X\SE(3)$ defines the left invariant metric on $\SE(3)$ as $\langle XU_1, XU_2\rangle_X=\langle \langle U_1, U_2 \rangle \rangle$ for all $X \in \SE(3)$ and $U_1, U_2 \in \se(3)$.
	The map  $\bar{\psi}:\mathbb{R}^{4\times 4} \rightarrow \mathbb{R}^6$ is defined as:
	\begin{align*} 
		\bar{\psi}(\mathbb{A})=\begin{bmatrix}
			\psi(A)\\
			\frac{1}{2}b
		\end{bmatrix}, \quad \forall \mathbb{A}=\begin{bmatrix}
			A&b\\
			c^{\top}&d
		\end{bmatrix} \in \mathbb{R}^{4\times 4}
	\end{align*}
	with $A\in \mathbb{R}^{3\times 3},b, c \in \mathbb{R}^3, d \in \mathbb{R}$. Then, one has the following identities:
	\begin{align*}
		\langle\langle \mathbb{A}, x^{\wedge} \rangle\rangle&=2x^{\top}\bar{\psi}(\mathbb{A})\\
		\bar{\psi}(X^{\top}(I_4-X)\mathbb{A})&=-\bar{\psi}((I_4-X^{-1})\mathbb{A}) 
	\end{align*}
	for all $\mathbb{A} \in \mathbb{R}^{4\times 4}, x \in \mathbb{R}^6, X\in \SE(3)$. Define the adjoint map $\Ad: \SE(3) \rightarrow \mathbb{R}^{6\times 6}$ and the adjoint operator $\ad: \mathbb{R}^6 \rightarrow \mathbb{R}^{6\times 6}$ as:
	$$\Ad_{X}= \begin{bmatrix}
		R&\mathbf{0}\\
		p^{\times}R&R
	\end{bmatrix}  ,
	\ad_{\xi}=  \begin{bmatrix}
		\omega^{\times}& \mathbf{0}\\
		v^{\times}& \omega^{\times} 
	\end{bmatrix}  $$
	for all $X=\mathcal{T}(R,p)\in \SE(3)$ and $\xi = [\omega\T ~ v\T]\T \in \mathbb{R}^6$. Then, one has the following useful identities:
		\begin{align*}
			\Ad_{X^{-1}}&=\Ad_{X}^{-1}\\
			Xx^{\wedge}X^{-1}&=(\Ad_{X}x)^{\wedge} \\
			\ad_x x&=\mathbf{0} 
		\end{align*}
	for all $X \in \SE(3), x\in \mathbb{R}^6$. 
	
	\subsection{Graph Theory}\label{sec:graph_theory}
	Consider a network comprising $N$ rigid body systems, where the information flow between the individual systems is modeled by a graph $\mathcal{G}=(\mathcal{V}, \mathcal{E})$, with $\mathcal{V}=\{1,\dots, N\}$ and $\mathcal{E} \subseteq \mathcal{V} \times \mathcal{V}$ denoting the vertex and edge sets of graph $\mathcal{G}$, respectively. In an undirected graph, an edge $(i,j)\in \mathcal{E}$ indicates that the $i$-th and $j$-th rigid bodies interact with each other, which means information can be exchanged bidirectionally between them. 
	For each edge between two rigid bodies, one arbitrarily assigns an index, a positive end, and a negative end. Let $\mathcal{M}:=\{1,\dots, M\}$ denote the set of graph edges, with $M$ denoting the total number of edges, and subsets $\mathcal{M}_i^+\subset\mathcal{M}$ and $\mathcal{M}_i^-\subset\mathcal{M}$ denote the set of edges for which node $i$ is the positive end and the negative end, respectively. Define $\mathcal{N}_i:=\{j\in \mathcal{V}: (i,j)\in \mathcal{E}\}$ as the set of connected neighbors of the $i$-th rigid body, and $\mathcal{M}_i :=\mathcal{M}_i^+ \cup \mathcal{M}_i^-$ denote the set of edges connected to the $i$-th rigid body. If an edge $(i,j)\in \mathcal{E}$, indexed by $k \in \mathcal{M}$, is oriented from $j$ to $i$, then $k \in \mathcal{M}_i^+ \cap \mathcal{M}_j^-$ otherwise $k \in \mathcal{M}_j^+ \cap \mathcal{M}_i^-$.
	The classical incidence matrix, having a row for each vertex and a column for each edge, is given by 
	\begin{align*}
		B_{ik}=\left\{
		\begin{array}{ll}
			+1 \quad  & k \in \mathcal{M}_i^+ \\
			-1 \quad  & k \in \mathcal{M}_i^-\\
			0 \quad & otherwise 
		\end{array}
		\right.
	\end{align*}
	for all $ i \in \mathcal{V},  k\in \mathcal{M}$. 
	
	\subsection{Hybrid Systems Framework}
	Consider a hybrid system given by 
	\begin{align}\label{hysys_def}
		\mathcal{H}\left\{
		\begin{array}{ll}\dot{x}~ \in F(x)& x \in C\\
			x^+\in G(x) & x\in D \end{array} \right.
	\end{align} 
	where $x\in \mathbb{R}^n$ denotes the state, $x^+$ denotes the state after a jump, the \emph{flow map} $F$, governing the continuous evolution of the state by a differential inclusion, the \emph{jump map} $G$, governing the discrete evolution of the state by a difference inclusion, the \emph{flow set} $C$, indicating where the flows are allowed, and the \emph{jump set} $D$, indicating where jumps are allowed. A solution $x$ to $\mathcal{H}$ is parameterized by $(t,j) \in \mathbb{R}_{\geq 0} \times \mathbb{N}$, where $t$ and $j$ denote the time passed in the flow set and the number of discrete jumps that have occurred. A subset $\dom x \subset \mathbb{R}_{\geq 0} \times \mathbb{N}$ is a \emph{hybrid time domain} if for every $(T,J) \in \dom x$, $\dom x \cap ([0,T] \times \{0,1, \dots, J\})$ is a compact hybrid domain of the form $ \cup_{j=0}^J([t_j,t_{j+1}] \times \{j\})$ with some finite sequence of times $0=t_0 \leq t_1 \leq t_2 \leq \dots \leq t_{J+1}$. 
	A solution $x$ to the hybrid system $\mathcal{H}$ is called \emph{maximal} if it cannot be extended by either flowing or jumping, \emph{complete} if its domain $\dom x$ is unbounded, and \emph{precompact} if it is complete and the closure of its range is compact.
	\section{Problem statement}\label{Problem Statement}
	Consider the heterogeneous MRBSs with a set of $N$ rigid bodies. Let $X_i=\mathcal{T}(R_i,p_i)\in \SE(3)$ denote the pose of the $i$-th rigid body system, and $\xi_i=[\omega_i^{\top}, v_i^{\top}]^{\top} \in \mathbb{R}^6$ denote the group velocity of the $i$-th rigid body system  expressed in the body-fixed frame. For each rigid body system, the fully actuated dynamics on $\SE(3)\times\mathbb{R}^6$ are given by \cite{bullo1999tracking}
	\begin{align}\label{eqn:kinematic}
		\begin{cases}
			\dot{X}_i&=X_i\xi_i^{\wedge} \\
			\mathbb{I}_i\dot{\xi}_i&=\ad_{\xi_{i}}^{\top}\mathbb{I}_i\xi_i+u_{i}
		\end{cases}
	\end{align} 
	where $\mathbb{I}_i=\blkdiag(J_i,m_iI_3)  \in \mathbb{R}^{6\times 6}$ denotes the mass-inertia matrix of the $i$-th rigid body with $J_i=J_i^{\top}>0$ and $m_i>0$, and $u_{i}=[\tau_i^{\top}, f_i^{\top}]^{\top} \in \mathbb{R}^6$ is the torque-force control input.  
	We assume that the relative pose between two rigid bodies $(i,j)\in \mathcal{E}$ is available if they are connected (through sensor measurements or communication). One introduces the following important assumption for interaction topology among rigid bodies, described by the graph $\mathcal{G}$:
	\begin{assum} \label{assum:1}
		The interaction graph $\mathcal{G}$ for system \eqref{eqn:kinematic} is assumed to be connected and acyclic.
	\end{assum}
	Note that Assumption \ref{assum:1} is commonly considered in the problem of attitude/pose synchronization in \cite{wang2012dual,boughellaba2026global,mayhew2012quaternion}. 
	For each edge $k\in  \mathcal{M}_i^+ \cap \mathcal{M}_j^- \subset \mathcal{M}$ connecting rigid bodies $(i,j)\in \mathcal{E}$, we define the relative pose between them as $\bar{X}_k=\mathcal{T}(\bar{R}_k,\bar{p}_k):=X_j^{-1}X_i$. Then, in view of \eqref{eqn:kinematic}, one has the following error dynamics:
	\begin{align}    \label{eqn:errdynamics}  
		\begin{cases}
			\dot{\bar{X}}_k&=\bar{X}_k\bar{\xi}_k^{\wedge} \\                   \mathbb{I}_i\dot{\xi}_i&=\ad_{\xi_{i}}^{\top}\mathbb{I}_i\xi_i+u_{i} 
		\end{cases}   
	\end{align} 
	where $\bar{\xi}_k:=\xi_i-\Ad_{\bar{X}_k}^{-1}\xi_j$ denotes the geometric relative group velocity error. 
	Define the time-varying incidence matrix $\bar{B}(t):=[\bar{B}_{ik}]_{\substack{1\leq i \leq N, 1\leq k \leq M}} \in \mathbb{R}^{6N\times 6M}$ with 
	\begin{align}\label{Bt_def}
		\bar{B}_{ik}=\left\{
		\begin{array}{ll} I_6&~~k\in \mathcal{M}_i^+\\
			-\Ad_{\bar{X}_k}^{-\top} &~~ k \in \mathcal{M}_i^{-}\\
			\mathbf{0}& ~~otherwise \end{array}\right. , \forall i \in \mathcal{V},  k\in \mathcal{M}.
	\end{align}
	
	Unlike the classical incidence matrix $B$ defined in Section \ref{sec:graph_theory}, the matrix $\bar{B}(t)$ defined in \eqref{Bt_def} is time-varying and depends on the relative pose $\bar{X}_k$ of each edge. The following lemma presents two useful properties of $\bar{B}(t)$ under Assumption \ref{assum:1}:      
	\begin{lem}\label{lem: prop_barB} 
		Consider the matrix $\bar{B}(t)$ defined in \eqref{Bt_def} for system \eqref{eqn:kinematic} under Assumption \ref{assum:1}. Let $\bar{\xi}=[\bar{\xi}_1^{\top}, \bar{\xi}_2^{\top},\dots, \bar{\xi}_M^{\top}]^{\top}
		\allowbreak \in \mathbb{R}^{6M}$, $y\in \mathbb{R}^{6M}$ and $\xi=[\xi_1^{\top}, \xi_2^{\top}, \dots, \xi_N^{\top}]^{\top}
		\in \mathbb{R}^{6N}$. Then the following statements hold:
		\begin{itemize}
			\item [1)] $\bar{\xi}=\bar{B}(t)^{\top}\xi$;
			\item [2)] $\bar{B}(t)$ is full of column rank for all $t \geq 0$, and $\bar{B}(t)y=\mathbf{0}$ implies $y=\mathbf{0}$ .
		\end{itemize}
	\end{lem}
	\begin{pf} 
		See Appendix \ref{sec:lem_HE}. 
	\end{pf} 
	
	Our objective of this work consists in designing distributed hybrid feedback laws under Assumption \ref{assum:1}, with and without velocity measurements for the system \eqref{eqn:errdynamics},  such that the states $(\bar{X}_k,\xi_i)$ globally asymptotically converge to $(I_4,\mathbf{0})$ for all $k \in \mathcal{M}$ and $i\in \mathcal{V}$.
	
	\section{Potential function design on $(\SE(3) \times \mathbb{R})^M$} \label{Potential function design}
	The construction of the potential function is critical for pose synchronization of MRBSs. In this section, a systematic procedure for constructing a general potential function on $(\SE(3) \times \mathbb{R})^M$ is provided.
	\subsection{Potential Function on $\SE(3)$ and $\SE(3)\times \mathbb{R}$}\label{IVA}

	Consider the following real-valued function $V:\SE(3) \to \mathbb{R}$:
	\begin{align}\label{V_PF}
		V(X):=\frac{1}{2}\tr((I_4-X)\mathbb{A}(I_4-X)^{\top})
	\end{align}
	where the weighting matrix $\mathbb{A}=\mathbb{A}\T \in \mathbb{R}^{4\times4}$ and its decomposition are given by
	\begin{align}\label{AA_def}
		\mathbb{A}:=\begin{bmatrix}
			A&b\\
			b^{\top}&d
		\end{bmatrix} =\begin{bmatrix}
			I_3&bd^{-1}  \\
			\mathbf{0}&1
		\end{bmatrix}
		\begin{bmatrix}
			W&0\\
			\mathbf{0}&d
		\end{bmatrix}
		\begin{bmatrix}
			I_3&\mathbf{0}\\
			b^{\top}d^{-1}&1
		\end{bmatrix} 
	\end{align}
	with $A \in \mathbb{R}^{3\times 3}, b\in \mathbb{R}^3, d\in \mathbb{R}_{>0}$ and $W:= A-bb^{\top}d^{-1}$ is positive definite.  
	Then, one can verify that $V(X) =\tr(W(I_3-R))+\frac{d}{2}\|p-(I_3-R)bd^{-1}\|^2$ is a potential function on $\SE(3)$ with respect to $I_4$, \ie, $V(X)>0 $ for all $X\in \SE(3)\setminus\{I_4\}$ and $V(X)=0$ if and only if $X = I_4$. The set of all the critical points of $V(X)$, denoted by $\Upsilon_V =\{X \in \SE(3): \nabla_{X}V(X)=\mathbf{0}\}$, is given by \cite{wang2019hybrid}
	\begin{align}
		\Upsilon_V &:= \{I_4\} \cup \{X=\mathcal{T}(R, p): R=\mathcal{R}_a(\pi, v),  \nonumber \\
		&\qquad  p=(I_3-R_a(\pi, v))bd^{-1}, v\in \mathcal{E}_v(W)\} \label{def:Upsilon_V}
	\end{align} 
	where $\mathcal{R}_a$ is defined in \eqref{Rodrigues}, and $\mathcal{E}_v(W)$ denotes the set of all unit eigenvectors of $W$, and the set of all the undesired critical points is denoted by $\Upsilon_V \setminus \{I_4\}$. Due to the existence of such undesired critical points of $V$ on $\SE(3)$,  AGAS is the strongest stability result that one can achieve, with time-invariant continuous feedback related to the potential function $V$.

	To achieve GAS for the problem of pose tracking of a single rigid body system, a modified potential function on $\SE(3) \times \mathbb{R}$ has been proposed in \cite{wang2022hybrid} by introducing the following transformation map $\Gamma_{\mathbb{A}}: \SE(3) \times \mathbb{R} \rightarrow \SE(3)$:
	\begin{align}\label{Gamma_def}
		\Gamma_{\mathbb{A}}(X, \theta)
		:=X \mathcal{T}_{u_c}(\theta), \quad \mathcal{T}_{u_c}(\theta)=\exp(\theta u_c^{\wedge})   
	\end{align}
	with a constant vector $u_c\in \mathbb{R}^6$ and a real-valued scalar variable $\theta \in \mathbb{R}$ governed by hybrid dynamics. Then, a modified potential function $U: \SE(3) \times \mathbb{R}  \rightarrow \mathbb{R}_{\geq 0}$ with respect to $\mathcal{A}_0:=(I_4,0)$ is defined as:
	\begin{align}\label{U_def}  
		U(X, \theta) :&= V(\Gamma_{\mathbb{A}}(X, \theta)) + \frac{\gamma}{2}\theta^2 \nonumber \\
		& = \frac{1}{2}\tr((I_4-\Gamma_{\mathbb{A}})\mathbb{A} (I_4-\Gamma_{\mathbb{A}})^{\top})+\frac{\gamma}{2}\theta^2
	\end{align}
	with some $\gamma>0$, $\mathbb{A}$ given by \eqref{AA_def} and $\Gamma_{\mathbb{A}}=\Gamma_{\mathbb{A}}(X, \theta)$ for the sake of simplicity. Using the fact $\Gamma_{\mathbb{A}}(X, 0) = X$, it is not difficult to verify from \eqref{V_PF} that $U(X,\theta)> 0$ for all $(X, \theta) \in (\SE(3) \times \mathbb{R}) \setminus \{\mathcal{A}_0\}$, and $U(X,\theta)= 0$ if and only if $(X, \theta)=\mathcal{A}_0$. Therefore, $U(X, \theta)$ in \eqref{U_def} is a potential function on $\SE(3) \times \mathbb{R}$ with respect to $\mathcal{A}_0$. Define the set of all the critical points of $U$ in \eqref{U_def} as $\Upsilon_U:=\{(X, \theta) \in \SE(3) \times \mathbb{R}: \nabla_{X}U(X, \theta) = \mathbf{0}, \nabla_{\theta}U(X, \theta)=0\}$. 
	The following lemma  provides some useful properties of the map $\Gamma_{\mathbb{A}}$ in \eqref{Gamma_def} and the potential function $U$ in \eqref{U_def}:
	
	\begin{lem}{\cite[Proposition 3]{wang2022hybrid}}\label{lem:Uproperty} 
		Consider the transformation map $\Gamma_{\mathbb{A}}$ defined in \eqref{Gamma_def}, the potential function $U$ defined in \eqref{U_def}, and the trajectories generated by $\dot{X}=X\xi^{\wedge}$ and $\dot{\theta}=\nu$ with $(X(0),\theta(0) )\in \SE(3)\times \mathbb{R}, (\xi(t), \nu(t)) \in  \mathbb{R}^6 \times \mathbb{R}, \forall t\geq 0$. Then, the following statements hold:
		\begin{subequations}\label{sts}
			\begin{align}
				\dot{\Gamma}_{\mathbb{A}}(X, \theta)&=\Gamma_{\mathbb{A}}(X, \theta) (\Ad^{-1}_{\mathcal{T}_{u_c}(\theta)}\xi+\nu u_c)^{\wedge}\label{st_a} \\
				\bar{\psi}_{\nabla}(X, \theta)&=\Ad^{-\top}_{\mathcal{T}_{u_c}(\theta)}\bar{\psi}((I_4-\Gamma_{\mathbb{A}}^{-1}(X, \theta))\mathbb{A}) \label{st_b}\\
				\nabla_{\theta}U(X, \theta)&=\gamma \theta+2u_c^{\top}\bar{\psi}((I_4-\Gamma_{\mathbb{A}}^{-1}(X, \theta))\mathbb{A}) \label{st_c}\\
				\mathcal{A}_0 &\in \Upsilon_U: = \Upsilon_V \times \{0\} \label{U_cri}
			\end{align}
		\end{subequations} 
		with $\bar{\psi}$ and $\Upsilon_V$ defined in Section 2.2 and \eqref{def:Upsilon_V}, respectively, $\bar{\psi}_{\nabla}(X, \theta) := \bar{\psi}(X^{-1}\nabla_{X}U(X, \theta))$  for the sake of simplicity. 
	\end{lem}
	By considering a block diagonal matrix $\mathbb{A}= \blkdiag(A,d)$ and $u_c = [u_{c1}\T, \mathbf{0}\T]\T$ with $u_{c1} \in \mathbb{S}^2$ related to the matrix $A$, the potential function $U$ defined in \eqref{U_def} can be simplified as $U(X, \theta)=\tr(W(I_3-R\mathcal{R}_a(\theta, u_{c1})))+\frac{d}{2}\|p\|^2+\frac{\gamma}{2}\theta^2.$
	Since the second position-only term in $U(X, \theta)$ is independent of the rotation, it can simplify the hybrid feedback control design, resulting in a strict decrease in the potential function over each jump, as shown in \cite{wang2022hybrid}. 
	Such an idea has been considered in \cite{hatanaka2012passivity} to deal with the pose synchronization problem with the weighting matrix $\mathbb{A}$ chosen as a block diagonal matrix $\mathbb{A}:= $diag$(\frac{1}{2}I_3,\frac{1}{\sqrt{2}})$.
	However, the choice of a block diagonal matrix $\mathbb{A}$ is very restrictive in practical applications. 
	For instance, in some applications such as in \cite{wang2019hybrid}, the weighting matrix $\mathbb{A}:= \sum_{i=1}^{n}k_ir_ir_i^{\top}$ is directly constructed from the inertial frame landmark positions $p_i \in \mathbb{R}^3$ with $r_i=[p_i^\top ~ 1]^{\top} \in \mathbb{R}^4$ and $k_i > 0$. In this situation, it is difficult to ensure that the matrix $\mathbb{A}$ is in a block diagonal form, due to the physical locations of the landmarks.
	
	\subsection{Generic Potential Function on $(\SE(3)\times \mathbb{R})^M$}
	To overcome the above-mentioned limitation, we aim to construct a generic potential function on $\SE(3) \times \mathbb{R}$ with a general matrix $\mathbb{A}$ defined in \eqref{AA_def}. 
	Let $u_c=[u_{c1}^{\top}, u_{c2}^{\top}]^{\top} \in \mathbb{R}^6$ with $u_{c1} \in \mathbb{S}^2$ and $u_{c2}\in \mathbb{R}^3$, one can rewrite $\mathcal{T}_{u_c}(\theta) =\exp(\theta u_c^{\wedge})$ as
	\begin{align}\label{eqn:exp_thetak}  
		\mathcal{T}_{u_c}(\theta) 
		= \begin{bmatrix}
			\exp(\theta u_{c1}^{\times}) & ~~~ \mathcal{U}(\theta u_{c1}^{\times}) \theta u_{c2}\\
			\mathbf{0}&1
		\end{bmatrix}
	\end{align}
	where the map $\mathcal{U}$ is defined as $\mathcal{U}(0 u_{c1}^{\times})= I_3$ when $\theta=0$ and 
	$ 
	\mathcal{U}(\theta u_{c1}^{\times}):=
	I_3+(\frac{1-\cos (\theta)}{\theta})u_{c1}^{\times}+(\frac{\theta-\sin (\theta)}{\theta})(u_{c1}^{\times})^2
	$ when $\theta\neq 0$. 
	In view of \eqref{AA_def}, \eqref{Gamma_def} and \eqref{eqn:exp_thetak}, the potential function $U(X, \theta)$ in \eqref{U_def} can be rewritten as:
	\begin{align} 
		&U(X,\theta)
		=\tr(W(I_3-R\mathcal{R}_a(\theta, u_{c1}))) +\frac{\gamma}{2}\theta^2 \nonumber\\
		&~~+\frac{d}{2}\|p+R\mathcal{U}(\theta u_{c1}^{\times})\theta u_{c2}-(I_3-R\mathcal{R}_a(\theta, u_{c1}))bd^{-1} \|^2 \nonumber \\
		&=\tr(W(I_3-R\mathcal{R}_a(\theta, u_{c1}))) +\frac{\gamma}{2}\theta^2 \nonumber\\
		&~~+\frac{d}{2}\|p-(I_3-R)bd^{-1}+R(\mathcal{R}_a(\theta,u_{c1})-I_3)bd^{-1} \nonumber\\
		&~~+R\mathcal{U}(\theta u_{c1}^{\times})\theta u_{c2}\|^2. \nonumber 
	\end{align}
	To simplify this potential function $U$, for a general weighting $\mathbb{A}$ defined in \eqref{AA_def}, we pick a constant vector $u_c:=[u_{c1}^{\top},u_{c2}^{\top}]^{\top} \in \mathbb{R}^6$ for the design of the transformation map $\Gamma_{\mathbb{A}}$ in \eqref{Gamma_def} with $u_{c1} \in \mathbb{S}^2$ and $u_{c2}=-u_{c1}^{\times} bd^{{-1}}$, where the unit vector $u_{c1}$ is related to the matrix $A$ to be given later. Then, using the fact $(u_{c1}^{\times})^3 = -\|u_{c1}\|^2 u_{c1}^{\times} = -u_{c1}^{\times}$ and the definition of $\mathcal{U}$ in \eqref{eqn:exp_thetak}, one can show that $-\mathcal{U}(\theta u_{c1}^{\times})\theta u_{c1}^\times  = I_3- \mathcal{R}_a(\theta ,u_{c1})$. Therefore, substituting this choice of $u_{c2}$ into $U(X, \theta)$, one has $R(\mathcal{R}_a(\theta,u_{c1})-I_3)bd^{-1} +R\mathcal{U}(\theta u_{c1}^{\times})\theta u_{c2}=\mathbf{0}$ and thus the potential function $U(X, \theta)$ can be simplified as: 
	\begin{align}\label{U_explicit}
		U(X, \theta)
		&=\tr(W(I_3-R\mathcal{R}_a(\theta, u_{c1})))+\frac{\gamma}{2}\theta^2 \nonumber\\
		&~~~~+\frac{d}{2}\|p - (I_3 - R)bd^{-1}\|^2. 
	\end{align}
	It follows from \eqref{U_explicit} that a change in $\theta$ will only affect the item $\tr(W(I_3-R\mathcal{R}_a(\theta, u_{c1})))+\frac{\gamma}{2}\theta^2$, while leaving the item $\frac{d}{2}\|p - (I_3 - R)bd^{-1}\|^2$ unchanged. This nice feature will simplify the design of our hybrid feedback in the next section to avoid all the undesired equilibrium points through jumps and achieve GAS for the hybrid closed-loop system. Note that a similar result was achieved in \cite{wang2022hybrid} only for a block diagonal matrix $\mathbb{A} = \blkdiag(A,d)$ with $u_{c2} = \mathbf{0}$, which is a special case of the potential function considered in this work (\ie, $b= \mathbf{0}$ and $u_{c2} = -u_{c1}^\times bd^{-1} = \mathbf{0}$).

	Now, we summarize the following lemma, adapted from \cite[Proposition 2]{wang2022hybrid}, that will be useful for our hybrid feedback control design.
	\begin{lem}   \label{lem:parameters}  
		Consider the potential function $U$ in \eqref{U_def}
		and the matrix $\mathbb A$ in \eqref{AA_def}.
		Let $(\lambda_i^W,v_i^W)$ denote the eigenpairs of $W$,
		with $0<\lambda_1^W\leq\lambda_2^W<\lambda_3^W$, and define $
		s_\lambda={\sum_{\imath=1}^3 \sum_{k\neq \imath} \lambda_{\imath}^W \lambda_k^W}, c_{\lambda}={\lambda_1^W \lambda_3^W}/{(\lambda_3^W - \lambda_1^W)}$. The set of parameters $\mathcal{P}_\mathbb{A}:= \{\Theta, u_{c1}, u_{c2}, \gamma, \delta\}$ is given by 
		\begin{align} \label{eqn:P_A}
			\mathcal{P}_\mathbb{A} =\left\{
			\begin{aligned}
				\Theta &= \{|\theta_{\jmath}| \in (0,\pi], \jmath=1,\dots,m\}  \\
				u_{c1} &=  \sum_{i=1}^3\alpha_i v_i^{W}, \quad  u_{c2} =-u_{c1}^{\times}bd^{-1}\\
				\gamma &<  \frac{4\Delta_W^*}{\pi^2},  0< \delta  <\big(\frac{4\Delta_W^*}{\pi^2}-\gamma\big)\frac{\theta_{\max}^2}{2}
			\end{aligned}
			\right.
		\end{align}
		where $\theta_{\max}= \sup\nolimits_{\theta_k'\in \Theta} |\theta_k'|$, the constant scalars $\alpha_i\in [0,1]$ with $\sum_{i=1}^3\alpha_i^2=1$ and $\Delta^*_W >0$ are chosen as per one of the following three cases:
		\begin{itemize} 
			\item  [1)] if $\lambda_1^W = \lambda_2^W$, $\alpha_3^2 = 1 - {\lambda_2^W}/{\lambda_3^W}, \alpha_i^2={\lambda_2^W}/{(2\lambda_3^W)}, i\in\{1,2\}$, and $\Delta_W^* = \lambda_1^W\bigl(1 - {\lambda_2^W}/{\lambda_3^W}\bigr)$;
			\item [2)] if $\lambda_2^W \geq c_{\lambda} $, $\alpha_1^2=0, \alpha_i^2 = {\lambda_i^W}/{(\lambda_2^W + \lambda_3^W)}, i\in\{2,3\}$, and $\Delta_W^* = \lambda_1^W$;
			\item [3)] if $\lambda_1^W < \lambda_2^W < c_{\lambda}$, $\alpha_i^2 = 1 - {4 \prod_{j\neq i} \lambda_j^W}/s_\lambda$, $i\in\{1,2,3\}~ \text{and} ~ \Delta_W^* = {4\lambda_1^W\lambda_2^W\lambda_3^W}/s_\lambda$.
		\end{itemize}
		Then, for every $(X, \theta) \in \Upsilon_U \setminus 
		\{\mathcal{A}_0\}$ one has
		\begin{equation}\label{mu_U_def}
			\mu_{U}(X, \theta):=U(X, \theta)-\min_{\theta' \in \Theta}U(X, \theta')>\delta.
		\end{equation}
	\end{lem}
	\begin{pf}
		See Appendix \ref{sec:lem_parameters}. 
	\end{pf}
		Lemma~\ref{lem:parameters} provides admissible parameters $\mathcal P_{\mathbb A}$ in \eqref{eqn:P_A} for constructing the potential function $U$.
		In particular, choosing $u_{c1}=\sum_{i=1}^3\alpha_i v_i^{W}$ according to \cite[Proposition 2]{berkane2017construction} and $u_{c2}=-u_{c1}^{\times}b/d$ guarantees the condition \eqref{mu_U_def} with a maximized gap $\delta$. 
		As discussed in \cite{mayhew2013global}, the positive threshold $\delta$ introduces hysteresis to suppress unnecessary jumps due to small measurement disturbances. Moreover, a singleton $\Theta$ is sufficient for the inequality in \eqref{mu_U_def}, and additional elements provide more candidates for the transformation map $\Gamma_A$ at each jump to minimize the potential function $U$, resulting in better performance of the system.
		
	Let $x:=(\bar{X}_1, \theta_1, \dots, \bar{X}_M, \theta_M) \in \mathcal{S}:=(\SE(3) \times \mathbb{R})^M$ denote the extended state of the MRBSs. Define the closed set $\mathcal{A}:=\{x\in \mathcal{S}: (\bar{X}_k, \theta_k)=\mathcal{A}_0, \forall k \in \mathcal{M}=\{1,\dots, M\}\}$. Then, from the potential function $U$ in \eqref{U_def}, we construct the following synergetic potential function $\bar{U}$ on $\mathcal{S}$ with respect to $\mathcal{A}$ for MRBSs: 
	\begin{align}\label{barU_def} 
		\bar{U}(x) =\sum_{k=1}^M U(\bar{X}_k, \theta_k)
	\end{align}
	whose gradients with respect to $\bar{X}_k$ and $\theta_k$ for each $k \in \mathcal{M}$ are given by $\nabla_{\bar{X}_k} \bar{U}(x)=\nabla_{\bar{X}_k} U(\bar{X}_k, \theta_k) $ and $\nabla_{\theta_k} \bar{U}(x)=\nabla_{\theta_k} U(\bar{X}_k, \theta_k)$.
	From \eqref{U_cri} and \eqref{barU_def}, 
	the set of all the critical points of $\bar{U}$ is given by $\Upsilon_{\bar{U}}:=\{x\in \mathcal{S}: (\bar{X}_k, \theta_k) \in \Upsilon_U, \forall k \in \mathcal{M}\}$.
	
	\section{Hybrid pose synchronization scheme with group velocity measurements}\label{Hybrid Pose Synchronization Scheme}
	Relying on the hybrid dynamical systems framework \citep{goebel2009hybrid}, we propose the following hybrid feedback control scheme for global pose synchronization of system \eqref{eqn:kinematic}:
	\begin{align}\label{hybridcontroller}
		&  u_{i}  = -k_X \sum\limits_{\kappa\in \mathcal{M}_i}\bar{B}_{i\kappa} \bar{\psi}_{\nabla}(\bar{X}_{\kappa}, \theta_{\kappa})
		-K_{\xi}\xi_i -K_e\sum\limits_{j \in \mathcal{N}_i}(\xi_i-\xi_j) \nonumber \\[2pt]
		&
		\left\{ \begin{aligned}
			&\dot{\theta}_k=f(\bar{X}_k, \theta_k), \quad ~(\bar{X}_k, \theta_k) \in \mathcal{F}^o_k \\
			&\theta_k^+ \in g(\bar{X}_k, \theta_k), \quad~  (\bar{X}_k, \theta_k) \in \mathcal{J}^o_k 
		\end{aligned}\right.
	\end{align}
	for all $i \in \mathcal{V}$ and $k \in \mathcal{M}_i^+ \subseteq \mathcal{M}$, with $k_X >0$, $K_{\xi}, K_e\in \mathbb{R}^{6\times 6}$ being positive definite, $\bar{B}$ defined in \eqref{Bt_def} and the gradient-based map $\bar{\psi}_{\nabla}(X, \theta) = \bar{\psi}(X^{-1}\nabla_{X}U(X, \theta))$ in Lemma \ref{lem:Uproperty}. 
	The flow and jump sets $\mathcal{F}^o_k, \mathcal{J}^o_k$ are defined by
	\begin{align}\label{set_Fk_J_k}
		\mathcal{F}_k^o :&=\{(\bar{X}_k, \theta_k) \in \SE(3) \times \mathbb{R}: \mu_{U}(\bar{X}_k, \theta_k) \leq \delta \} \nonumber \\ 
		\mathcal{J}_k^o:&= \{(\bar{X}_k, \theta_k) \in \SE(3) \times \mathbb{R}: \mu_{U}(\bar{X}_k, \theta_k) \geq \delta\}
	\end{align}
	with the map $\mu_{U}$ and the constant $\delta$ given by Lemma \ref{lem:parameters}. The flow map $f: \SE(3) \times \mathbb{R} \rightarrow  \mathbb{R}$ and the jump map $g: \SE(3) \times \mathbb{R} \rightrightarrows \Theta$ are given by
	\begin{align}\label{eqn: gtheta}
		f(\bar{X}_k, \theta_k):&= -k_{\theta}\nabla_{\theta_k}U(\bar{X}_k,\theta_k), \quad k_{\theta}>0 \nonumber \\
		g(\bar{X}_k, \theta_k):&= \arg\min_{\theta_k' \in \Theta}U(\bar{X}_k, \theta'_k).
	\end{align}
	By \eqref{set_Fk_J_k} and \eqref{mu_U_def}, all the undesired critical points of $U$ lie in $\mathcal J_k^o$ and are excluded from $\mathcal F_k^o$. The switching variable $\theta_k$ follows the gradient-based flow map $f$ on $\mathcal F_k^o$ (\ie, away from the undesired critical set), and resets through the jump map $g$ to a nonzero candidate in $\Theta$ on $\mathcal J_k^o$ (\ie, close to the undesired critical set). The control input \eqref{hybridcontroller} combines gradient-based relative pose feedback for synchronization, velocity damping to drive $\xi_i$ to zero, and relative velocity feedback that may improve transient performance. 
	
	The map $\Gamma_A$ in \eqref{Gamma_def} transforms the relative pose $\bar X_k$ into $\bar X_k' =\bar X_k\mathcal T_{u_c}(\theta_k)$. The parameter choice $\mathcal P_{\mathbb A}$ in \eqref{eqn:P_A} ensures that, for each undesired critical point $(\bar X_k,\theta_k)$, there exists a new point $(\bar{X}_k\mathcal{T}_{u_c}(\theta'),\theta')$ with $\theta'\in \Theta$ yielding a lower value in potential function $U$ as in \eqref{mu_U_def}. Hence, resetting $\theta_k$ to $\theta_k^+\in\arg\min_{\theta_k'\in\Theta} U(\bar X_k,\theta_k')$
	changes the transformed state to $(\bar X_k\mathcal T_{u_c}(\theta_k^+),\theta_k^+)$ and decreases $U$ by at least $\delta$. The gradient-based control input $u_i$ in \eqref{hybridcontroller} and the flow map $f$ in \eqref{eqn: gtheta} drive the state toward the critical set $\Upsilon_U=\Upsilon_V\times\{0\}$, where the gradients vanish. The jump map $g$ in \eqref{eqn: gtheta} and the jump set in \eqref{set_Fk_J_k} allow the transformed state to jump away from undesired critical points and restore nonzero gradients. For the multi-body system, consider the extended
	potential function $\bar U(x)=\sum_{k=1}^M U(\bar X_k,\theta_k)$ in \eqref{barU_def}, defined on $(\SE(3)\times\mathbb R)^M$, where $M$ is the number
	of edges. As it to be shown in the proof of Theorem~\ref{T1}, the Lyapunov function
	$k_X\bar U(x)+\sum_{i=1}^N \xi_i^\top\mathbb I_i\xi_i$ is nonincreasing along flows and decreases by at least $k_X\delta$ at every jump, which establishes GAS of the equilibrium set $\mathcal{A}\times\{\mathbf0\}$ and finiteness of the number of jumps.

	Define the new states $x_c:=(x, \xi) \in \mathcal{S}_c:=(\SE(3) \times \mathbb{R})^M \times \mathbb{R}^{6N}$ and $\xi: =[\xi_1^{\top},\dots,\xi_N^{\top}]^{\top} \in \mathbb{R}^{6N}$.
	Define $\bar{\mathcal{A}}:= \mathcal{A} \times \{\mathbf{0}\}=\{x_c \in \mathcal{S}_c:x\in \mathcal{A}, \xi=\mathbf{0} \}=\{x_c \in \mathcal{S}_c: (\bar{X}_k, \theta_k)=(I_4,0), \xi_i=\mathbf{0} , \forall k \in \mathcal{M}, i\in \mathcal{V}\}$, which is a singleton and therefore compact. 
	From \eqref{eqn:errdynamics} and \eqref{hybridcontroller}, one obtains the following hybrid closed-loop systems:
	\begin{align}\label{hybsys} 
		\left\{  
		\begin{array}{ll} 
			\dot{x}_c=\begin{pmatrix}
				F_x(x_c)\\
				F_{\xi}(x_c)
			\end{pmatrix}, &x_c \in \mathcal{F}_c:=  \mathcal{F}\times \mathbb{R}^{6N}\\
			x_c^+ \in G(x_c), &x_c \in \mathcal{J}_c:=\mathcal{J}\times \mathbb{R}^{6N}
		\end{array}    \right. 
	\end{align}
	where $\mathcal{F}:= \{x\in \mathcal{S}:~ \forall k \in \mathcal{M}_i^+, i \in \mathcal{V},~ (\bar{X}_k, \theta_k) \in \mathcal{F}^o_k\}$ and $\mathcal{J}:=\{x\in \mathcal{S}:~ \exists k \in \mathcal{M}_i^+, i \in \mathcal{V},~ (\bar{X}_k, \theta_k) \in \mathcal{J}^o_k\}$, 
	\begin{align*}  \begin{smallmatrix}
			& F_x(x_c)&:=\begin{pmatrix}  \begin{smallmatrix}
					\bar{X}_1 \bar{\xi}_1^{\wedge} \\
					f(\bar{X}_1, \theta_1)\\
					\vdots\\
					\bar{X}_M \bar{\xi}_M^{\wedge} \\
					f(\bar{X}_M, \theta_M)\\
			\end{smallmatrix} \end{pmatrix} , ~~~
			F_{\xi}(x_c):=\begin{pmatrix} \begin{smallmatrix}
					\mathbb{I}_1^{-1}(\ad^{\top}_{\xi_1}\mathbb{I}_1\xi_1+u_{1})\\
					\vdots\\
					\mathbb{I}_N^{-1}(\ad^{\top}_{\xi_N}\mathbb{I}_N\xi_N+u_{N}) \end{smallmatrix}
			\end{pmatrix}  \\
		\end{smallmatrix}
	\end{align*}
	and $G(x_c) :=(\bar{X}_1, g_{\theta}(\bar{X}_1, \theta_1), \dots, \bar{X}_M, g_{\theta}(\bar{X}_M, \theta_M), \xi_1, \allowbreak \dots, \xi_N)$,
	$f$ and $g$ defined in \eqref{eqn: gtheta} and $u_i$  defined in \eqref{hybridcontroller}. The function $g_{\theta}$ is defined as: $g_{\theta}(\bar{X}_k, \theta_k)=g(\bar{X}_k, \theta_k)$ if $(\bar{X}_k, \theta_k)\in \mathcal{J}_k^o$, otherwise $g_{\theta}(\bar{X}_k, \theta_k)=\theta_k$.
	Note that $\bar{\xi}_k=\xi_i-\Ad_{\bar{X}_k}^{-1} \xi_j$ as defined  after \eqref{eqn:errdynamics}. 
	Hence, the closed-loop dynamics in \eqref{hybsys} depend only on $x_c$,  and the system \eqref{hybsys} is autonomous.
	The following lemma shows that the hybrid closed-loop system \eqref{hybsys} satisfies the hybrid basic conditions \citep{goebel2009hybrid}, guaranteeing the well-posedness of the hybrid closed-loop system:   
	\begin{lem}\label{lem:hybrid basic condition}   
		The hybrid closed-loop system \eqref{hybsys} satisfies the following conditions:  
		\begin{itemize}
			\item [$A1$)] $\mathcal{F}_c$ and $\mathcal{J}_c$ are closed subsets of $\mathbb{R}^n$;\\[-0.1in]
			\item [$A2$)] $F(x_c):=(F_x(x_c), F_{\xi}(x_c))$ is outer semicontinuous and locally bounded relative to $\mathcal{F}_c, \mathcal{F}_c \subseteq \dom F$, and $F(x_c)$ is convex for every $x_c \in \mathcal{F}_c$;\\[-0.1in]
			\item [$A3$)] $G$ is outer semicontinuous and locally bounded relative to $\mathcal{J}_c, \mathcal{J}_c \subseteq \dom G$.\\[-0.1in]
		\end{itemize} 
	\end{lem}
	\begin{pf}
		See Appendix~\ref{sec:lem_hybrid_basic_condition}.
	\end{pf}

	Now, one can state the following main result:
	\begin{thm}\label{T1}
		Consider the hybrid closed-loop system \eqref{hybsys} with $f$ and $g$ in \eqref{eqn: gtheta} and $u_i$ in \eqref{hybridcontroller}. Let $k_X, k_{\theta}>0, K_{\xi}, K_e \in \mathbb{R}^{6 \times 6}$ be positive definite, and suppose that Assumption \ref{assum:1} holds and the set $\mathcal{P}_\mathbb{A}$ is chosen as per Lemma \ref{lem:parameters}. Then, the closed set $\bar{\mathcal{A}}$ is globally asymptotically stable for the hybrid closed-loop system \eqref{hybsys} and the number of jumps is finite.
	\end{thm}
	\begin{pf}
		See Appendix~\ref{sec:pfT1}.
	\end{pf}
	
	\begin{rem}
		The proof of Theorem~\ref{T1} relies on the preceding lemmas. Specifically, Lemma 4 verifies that the closed-loop dynamics \eqref{hybsys} satisfy the hybrid basic conditions required for the stability analysis. Lemma \ref{lem:Uproperty} provides the gradient and critical point expressions of $U$ used in the hybrid feedback to establish that the Lyapunov function is nonincreasing along flows, while Lemma \ref{lem:parameters} specifies parameters to ensure a strict decrease in the Lyapunov function at each jump. Moreover, the full-column-rank property in Lemma \ref{lem: prop_barB} is used in the invariance argument to establish convergence of the gradients of $U$ to zero. Therefore, based on the hybrid switching mechanism, the result of GAS for the equilibrium set $\bar{\mathcal{A}}$ of the closed-loop system \eqref{hybsys} is guaranteed.
	\end{rem}

\begin{figure}[!ht]
		\centering
		\includegraphics[width=0.45\textwidth]{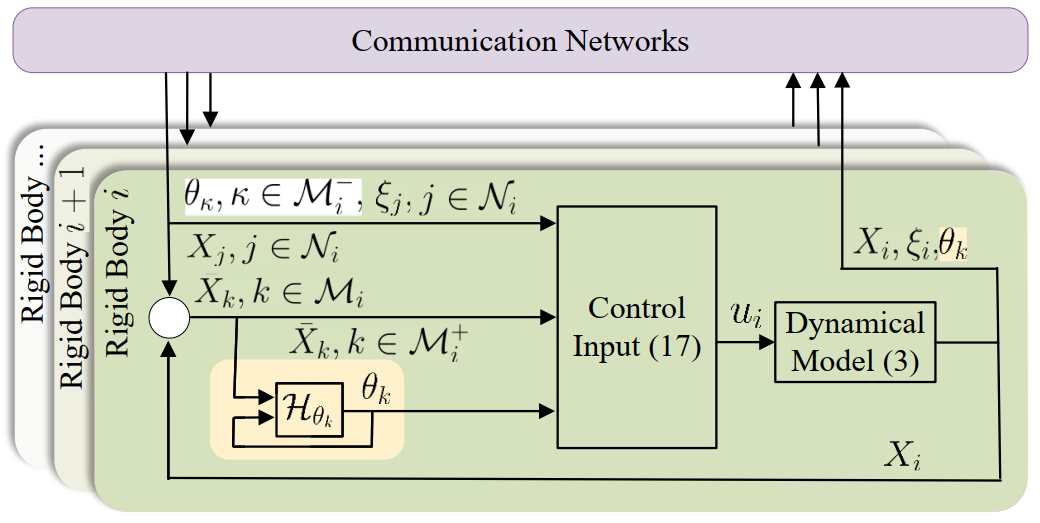}
		\caption{Block diagram of the distributed hybrid feedback for pose synchronization: a) if $\mathcal{M}_i^+ \neq\emptyset$ and $\mathcal{M}_i^- =\emptyset$, the $i$-th rigid body uses only the locally implemented $\theta_k, \forall k \in \mathcal{M}_i^+$; b) if $\mathcal{M}_i^+ =\emptyset$ and $\mathcal{M}_i^- \neq\emptyset$, the $i$-th rigid body uses only the received $\theta_\kappa$ from it's neighbors for all $\kappa \in \mathcal{M}_i^-$; c) if $\mathcal{M}_i^+ \neq\emptyset$ and $\mathcal{M}_i^- \neq\emptyset$, the $i$-th rigid body uses both the locally implemented $\theta_k, \forall k \in \mathcal{M}_i^+$ and the received $\theta_\kappa, \forall \kappa \in \mathcal{M}_i^-$.} 
		\label{fig:block diagram1}
	\end{figure}

	From the definition of $\bar{B}$ in \eqref{Bt_def}, the gradient-based relative pose feedback term in the control input $u_i$ in \eqref{hybridcontroller} can be decomposed as $\sum_{\kappa\in \mathcal{M}_i}\bar{B}_{i\kappa} \bar{\psi}_{\nabla}(\bar{X}_{\kappa}, \theta_{\kappa}) = \sum_{q \in \mathcal{M}_i^+} \bar{\psi}_{\nabla}(\bar{X}_{q}, \theta_{q}) -\sum_{l \in \mathcal{M}_i^-}\Ad_{\bar{X}_l}^{-\top} \bar{\psi}_{\nabla}(\bar{X}_{l}, \theta_{l})$. 
	In view of the potential function $U$ constructed in \eqref{U_def}, \eqref{st_b} and \eqref{st_c}, hybrid feedback law \eqref{hybridcontroller} can be explicitly implemented in the following form:
	\begin{align}
		&u_{i}  =-k_X \left(\sum\nolimits_{q \in \mathcal{M}_i^+} \Ad^{-\top}_{\mathcal{T}_{u_c}(\theta_q)}\bar{\psi}((I_4-\mathcal{T}_{u_c}^{-1}(\theta_q)\bar{X}_q^{-1})\mathbb{A}) \right. \nonumber \\
		&~~  \left. \quad-\sum\nolimits_{l \in \mathcal{M}_i^-}\Ad_{\bar{X}_ l}^{-\top} \Ad^{-\top}_{\mathcal{T}_{u_c}(\theta_ l)}\bar{\psi}((I_4- \mathcal{T}_{u_c}^{-1}(\theta_ l)\bar{X}_l^{-1})\mathbb{A}) \right) \nonumber \\
		& ~~  \quad -K_{\xi}\xi_i -K_e\sum\nolimits_{j \in \mathcal{N}_i}(\xi_i-\xi_j)\nonumber 
	\end{align}
	In practice, the hybrid dynamics of the auxiliary variable $\theta_k$ associated with edge $k\in\mathcal{M}_i$ are implemented by the $i$-th rigid body if the edge satisfies $k\in\mathcal{M}_i^+$, otherwise, the value of $\theta_k$ is received by the $i$-th rigid body from one of its connected neighbors if $k\in\mathcal{M}_i^-$.  
	Therefore, the proposed hybrid control law \eqref{hybridcontroller} can be implemented in a fully distributed manner (as shown in Fig. \ref{fig:block diagram1}). 

	\section{Hybrid pose synchronization without group velocity measurements}\label{velocityfree}
	The implementation of the distributed hybrid feedback control in the previous section requires the measurements of the group velocity for each rigid body. However, in practical scenarios, acquiring accurate velocity information is often challenging due to limited onboard payload, high implementation costs, significant measurement noise or potential sensor failures. To address this challenge, inspired by \cite{abdessameud2010global} and \cite{boughellaba2026global}, we develop a velocity-free hybrid pose synchronization scheme in this section. Specifically, to compensate for the absence of group velocity measurements, a hybrid auxiliary system with variables $(\hat{X}_i, \tilde{\theta}_i)$ on $\SE(3) \times \mathbb{R}$ is associated with each rigid body. Defining the error state $\tilde{X}_i:=\hat{X}_i^{-1}X_i\in \SE(3)$, we propose the following hybrid auxiliary system on $\SE(3) \times \mathbb{R}$:
	\begin{align}\label{dyncheckX}
		&~ \dot{\hat{X}}_i=\hat{X}_i( Ad_{\tilde{X}_i} \beta_i)^{\wedge}, \quad \beta_i = K_{\hat{X}} \bar{\psi}_{\nabla}(\tilde{X}_i, \tilde{\theta}_i) \nonumber \\
		& \left\{ \begin{aligned}
			&\dot{\tilde{\theta}}_i =f(\tilde{X}_i, \tilde{\theta}_i), \quad ~ (\tilde{X}_i,\tilde{\theta}_i)\in \mathcal{F}^o_i   \\
			&\tilde{\theta}_i^{+} \in g(\tilde{X}_i, \tilde{\theta}_i), \quad~  (\tilde{X}_i,\tilde{\theta}_i)\in \mathcal{J}^o_i
		\end{aligned} \right.
	\end{align}
	where $ K_{\hat{X}} \in \mathbb{R}^{6\times 6}$  is a positive definite matrix, $f$ and $g$ are defined in \eqref{eqn: gtheta}, the flow and jump sets $\mathcal{F}^o_i$ and $\mathcal{J}^o_i$ are given by 
	\begin{align*} 
		\mathcal{F}_i^o :&=\{(\tilde{X}_i, \tilde{\theta}_i) \in \SE(3) \times \mathbb{R}: \mu_{U}(\tilde{X}_i, \tilde{\theta}_i) \leq \delta \} \nonumber \\ 
		\mathcal{J}_i^o:&= \{(\tilde{X}_i, \tilde{\theta}_i) \in \SE(3) \times \mathbb{R}: \mu_{U}(\tilde{X}_i, \tilde{\theta}_i) \geq \delta\}.
	\end{align*}
	It follows from \eqref{eqn:kinematic} and \eqref{dyncheckX} that the dynamics of $\tilde{X}_i$ are given by
	\begin{align}\label{deqn:widetildeXdynamics}
		\dot{\tilde{X}}_i=\tilde{X}_i(\xi_i-\beta_i)^{\wedge}
	\end{align}
	for all $(\tilde{X}_i,\tilde{\theta}_i)\in \mathcal{F}^o_i$, and $\tilde{X}_i^{+}=\tilde{X}_i $ for all $(\tilde{X}_i,\tilde{\theta}_i)\in \mathcal{J}^o_i$. Define the extended state $\tilde{x}_c:=(x_c,\tilde{X}_1, \tilde{\theta}_1, \dots, \tilde{X}_N,  \tilde{\theta}_N)  \in \tilde{\mathcal{S}}_c := \mathcal{S}_c \times (\SE(3) \times \mathbb{R})^N$. Based on the auxiliary system \eqref{dyncheckX} and the potential functions $U(\bar{X}_k, \theta_k)$ and $U(\tilde{X}_i, \tilde{\theta}_i)$ constructed as \eqref{U_def}, we propose the following velocity-free hybrid control law:
	\begin{align}\label{velocityfreecontroller}
		u_{i}   = -k_X \sum\limits_{\kappa \in \mathcal{M}_i}\bar{B}_{i\kappa} \bar{\psi}_{\nabla}(\bar{X}_{\kappa}, \theta_{\kappa}) -k_{\tilde{X}}\bar{\psi}_{\nabla}(\tilde{X}_i, \tilde{\theta}_i) 
	\end{align}
	where $\theta_k$ and $\tilde{\theta}_i$ are given by \eqref{hybridcontroller} and \eqref{dyncheckX} for all $i \in \mathcal{V}, k\in \mathcal{M}_i^+$,  $k_X, k_{\tilde{X}}>0$.
	Unlike the hybrid feedback control law~\eqref{hybridcontroller}, which uses group velocity feedback for damping, the proposed scheme~\eqref{velocityfreecontroller} generates damping through the auxiliary system defined by ~\eqref{dyncheckX}, thereby removing the need for group velocity measurements.  From \eqref{eqn:errdynamics}, \eqref{deqn:widetildeXdynamics} and \eqref{velocityfreecontroller}, one has the following hybrid closed-loop dynamics: 
	\begin{align}\label{eqn:vlofreehybsys}
		\left\{
		\begin{array}{ll} 
			\dot{\tilde{x}}_c=\tilde{F}(\tilde{x}_c),\quad &\tilde{x}_c \in \tilde{\mathcal{F}}_c\\ [2pt]
			\tilde{x}_c^+ \in \tilde{G}(\tilde{x}_c), \quad &\tilde{x}_c \in \tilde{\mathcal{J}}_c
		\end{array} \right.
	\end{align}
	where $\tilde{\mathcal{F}}_c:=\{\tilde{x}_c \in \tilde{\mathcal{S}}_c: x_c \in \mathcal{F}_c \enspace {\rm and}  \enspace \forall i \in \mathcal{V}, (\tilde{X}_i, \tilde{\theta}_i) \in  \mathcal{F}^o_i \}$ and $\tilde{\mathcal{J}}_c:=\{\tilde{x}_c \in \tilde{\mathcal{S}}_c: x_c \in \mathcal{J}_c  \enspace {\rm or} \enspace \exists i \in \mathcal{V}, (\tilde{X}_i, \tilde{\theta}_i) \in  \mathcal{J}^o_i \}$ with $(\mathcal{F}_c, \mathcal{J}_c)$ defined in \eqref{hybsys} and $(\mathcal{F}^o_i,\mathcal{J}^o_i)$ defined after \eqref{dyncheckX}, the flow and jump maps are given by
	\begin{align*} \begin{smallmatrix}
			\tilde{F}(\tilde{x}_c):&=\left( \begin{smallmatrix}
				F_x(x_c)\\
				\mathbb{I}_1^{-1}(\ad^{\top}_{\xi_1}\mathbb{I}_1\xi_1+u_{1})\\
				\vdots\\
				\mathbb{I}_N^{-1}(\ad^{\top}_{\xi_N}\mathbb{I}_N\xi_N+u_{N})\\
				\tilde{X}_1(\xi_1-\beta_1)^{\wedge}\\
				f(\tilde{X}_1, \tilde{\theta}_1)\\
				\vdots\\
				\tilde{X}_N(\xi_N-\beta_N)^{\wedge}\\
				f(\tilde{X}_N, \tilde{\theta}_N)
			\end{smallmatrix} \right), ~
			\tilde{G}(\tilde{x}_c):&=\left( \begin{smallmatrix}
				G(x_c) \\
				\tilde{X}_1 \\
				g_{\tilde{\theta}}(\tilde{X}_1,\tilde{\theta}_1)\\
				\vdots \\
				\tilde{X}_N \\
				g_{\tilde{\theta}}(\tilde{X}_N, \tilde{\theta}_N))
			\end{smallmatrix} \right) 
		\end{smallmatrix}
	\end{align*}
	with $G(x_c), F_x(x_c)$ given after \eqref{hybsys}, $u_i$ given in \eqref{velocityfreecontroller}, $\beta_i$ defined in \eqref{dyncheckX}, $f$ and $g$ given in \eqref{eqn: gtheta}. 
	The function $g_{\tilde{\theta}}$ is defined as: $g_{\tilde{\theta}}(\tilde{X}_i, \tilde{\theta}_i)=g(\tilde{X}_i, \tilde{\theta}_i)$ if $(\tilde{X}_i, \tilde{\theta}_i) \in \mathcal{J}_i^o$, otherwise $g_{\tilde{\theta}}(\tilde{X}_i, \tilde{\theta}_i)=\tilde{\theta}_i$. One can verify that $\tilde{\mathcal{F}}_c\cup \tilde{\mathcal{J}}_c= \tilde{\mathcal{S}}_c$, $\tilde{\mathcal{F}}_c$ and $\tilde{\mathcal{J}}_c$ are closed, and the hybrid  system ~\eqref{eqn:vlofreehybsys} is autonomous and satisfies the hybrid basic conditions \citep{goebel2009hybrid}.
	
	Define the closed set $\tilde{\mathcal{A}}:
	=\{\tilde{x}_c \in \tilde{\mathcal{S}}_c: (\bar{X}_k, \theta_k)=\mathcal{A}_0, \xi_i=\mathbf{0}, (\tilde{X}_i, \tilde{\theta}_i)=\mathcal{A}_0,  \forall k \in \mathcal{M}, i\in \mathcal{V}\}$.  Then, one can state the following main result:
	\begin{thm}\label{T2}
		Consider the closed-loop system \eqref{eqn:vlofreehybsys} with $f, g$ defined in \eqref{eqn: gtheta}, and $\beta_i, i \in \mathcal{V}$ defined in \eqref{dyncheckX}. Suppose that the Assumption \ref{assum:1} holds, and choose $k_X, k_{\tilde{X}}, k_{\theta}>0$, $K_{\hat{X}}$ being positive definite and the set of parameters $\mathcal{P}_{\mathbb{A}}$ as per Lemma ~\ref{lem:parameters}. Then, the closed set $\tilde{\mathcal{A}}$ is globally asymptotically stable for the hybrid closed-loop system \eqref{eqn:vlofreehybsys} and the number of jumps is finite.
	\end{thm}
	\begin{pf}
		See Appendix~\ref{sec:pfT2}.
	\end{pf}

\begin{figure}[!ht]
		\centering
		\includegraphics[width=0.45\textwidth]{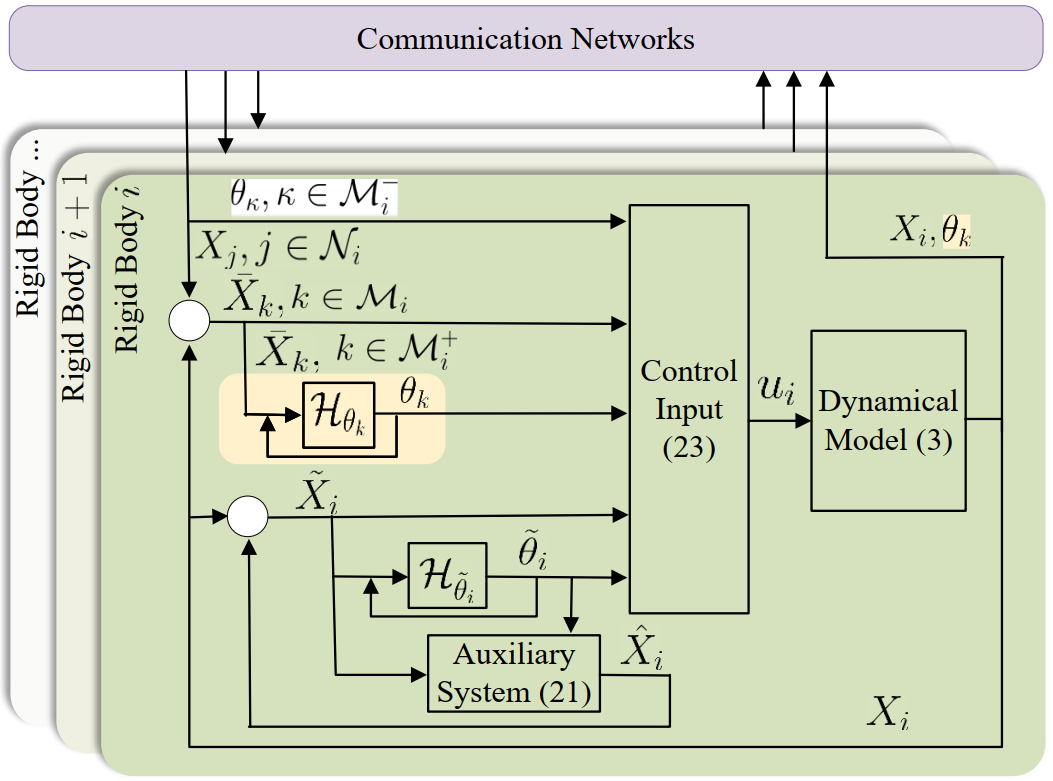}
		\caption{Block diagram of the distributed velocity-free hybrid feedback for pose synchronization. The hybrid dynamics of the auxiliary variables $\tilde{\theta}_i$ are only implemented by the $i$-th rigid body. The hybrid dynamics of the auxiliary variables $\theta_k$ are only implemented by the $i$-th rigid body when $k\in \mathcal{M}_i^+$, and the values of the auxiliary variables $\theta_{\kappa}$ are received by the $i$-th rigid body from its connected neighbors when $\kappa \in \mathcal{M}_i^-$.}  
		\label{fig:velocityfree block diagram}
	\end{figure}

	By following the same arguments as in Section \ref{Hybrid Pose Synchronization Scheme}, the proposed velocity-free hybrid control law \eqref{velocityfreecontroller} can be explicitly written in a similar form of $u_i$ in the end of Section~\ref{Hybrid Pose Synchronization Scheme} such that it can be implemented in a fully distributed manner. The diagram for the implementation of the proposed distributed hybrid feedback for pose synchronization without velocity measurements is shown in Fig. \ref{fig:velocityfree block diagram}.
	
	\section{Simulation}\label{simulation}
\begin{figure}[!ht]
		\centering
		\includegraphics[width=0.48\textwidth]{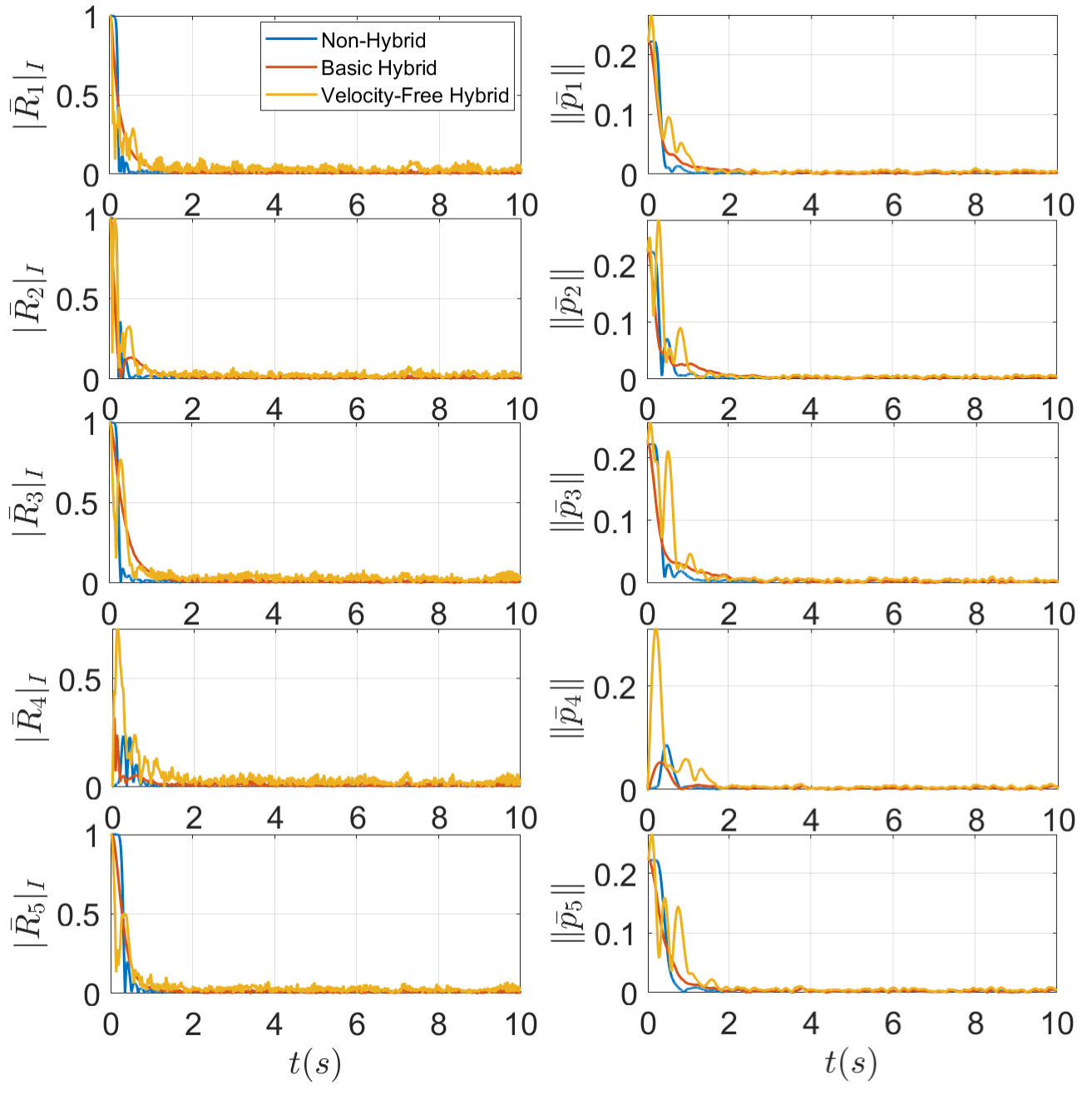}
		\vspace{-0.2in}
		\caption{Time evolution of the relative attitude and position associated with each edge with noisy measurements.} \vspace{0.1in}
		\label{simulation results4} 
		\centering
		\includegraphics[width=0.48\textwidth]{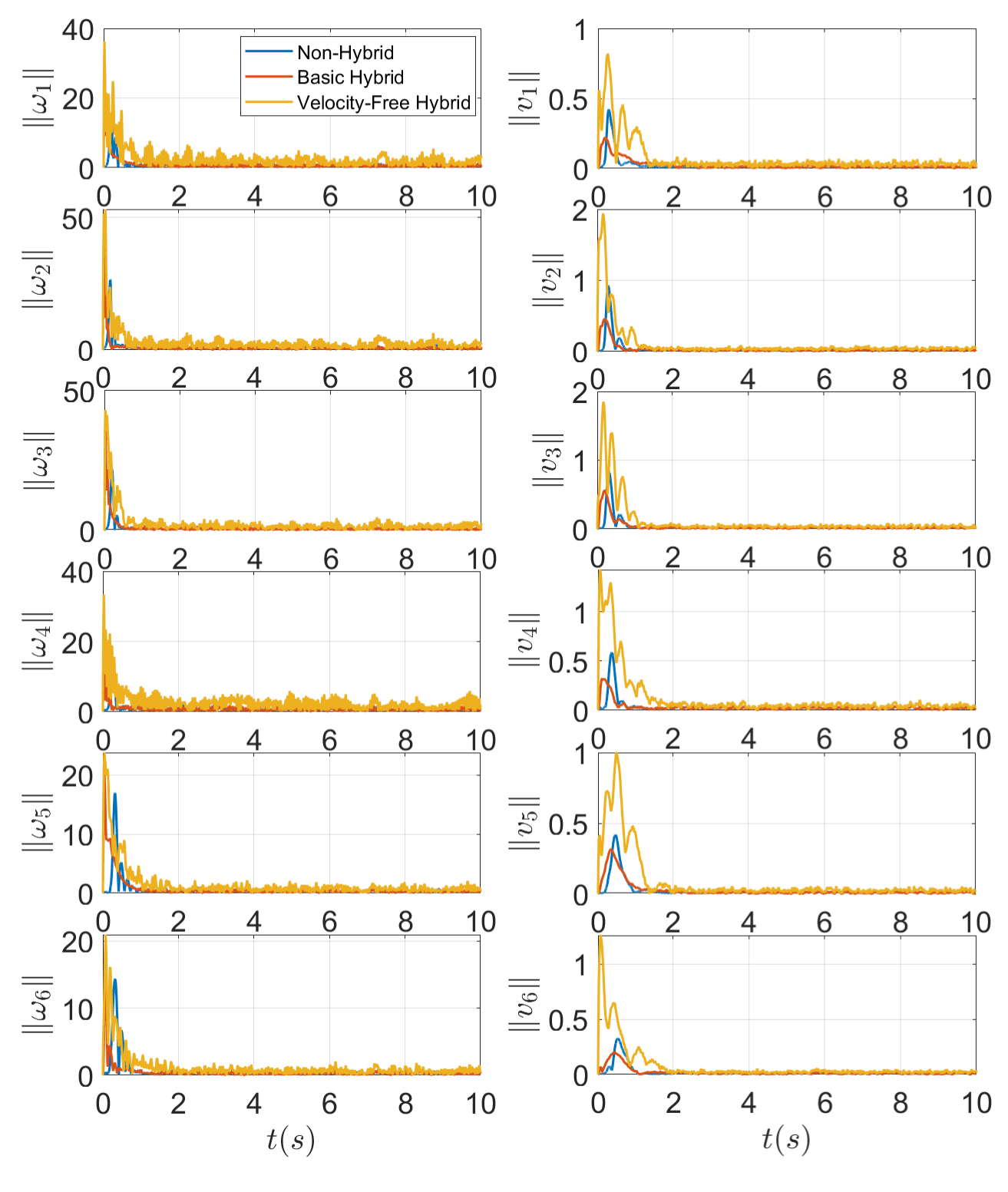}
		\vspace{-0.28in}
		\caption{Time evolution of the angular and linear velocity associated with each rigid body with noisy measurements.} 
		\label{simulation results5} 
	\end{figure}

\begin{figure}[!ht]
		\centering
		\includegraphics[width=0.48\textwidth]{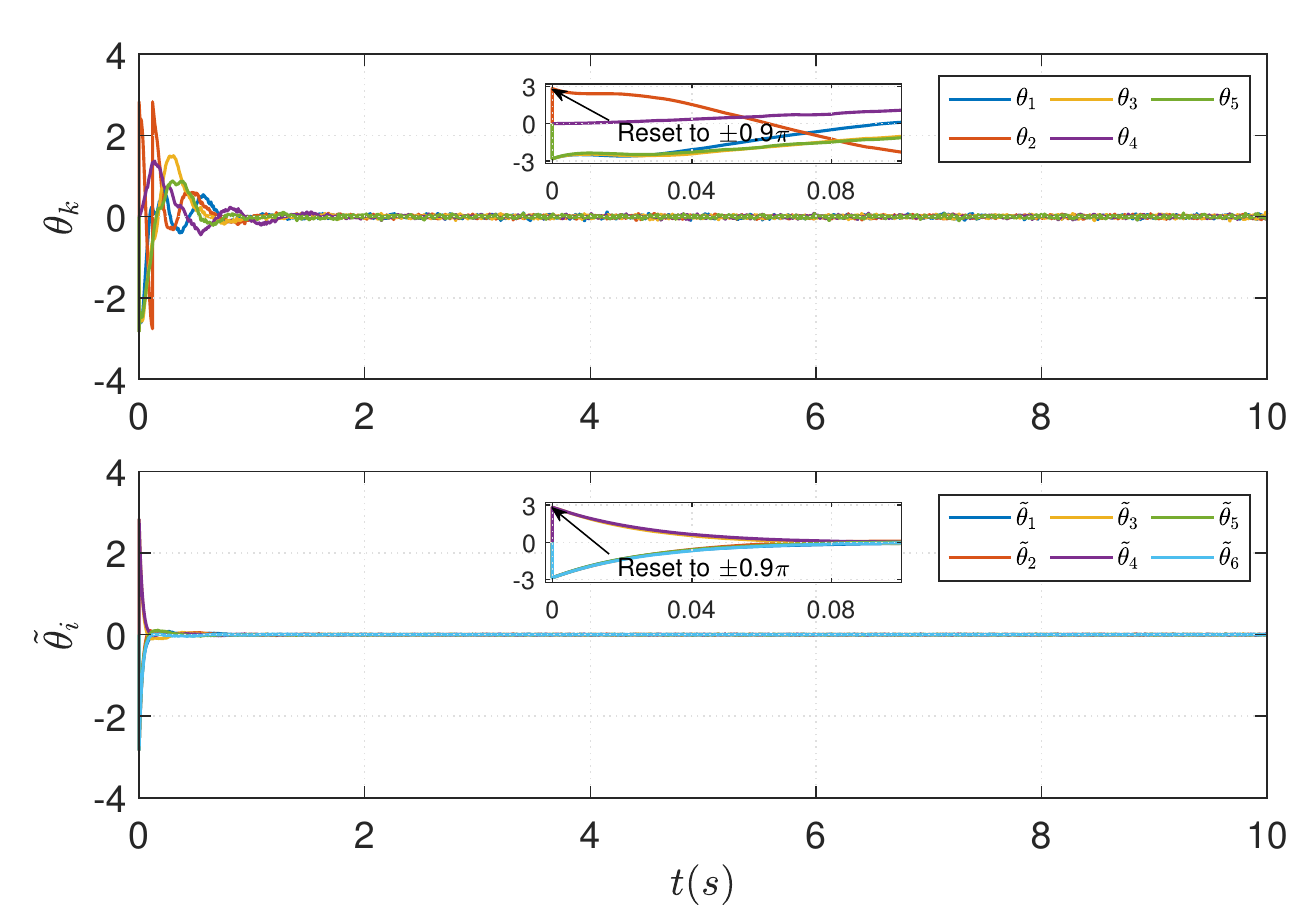}
		\vspace{-0.3in}
		\caption{Time evolution of the switching variables $\theta_k$ and $\tilde{\theta}_i$ with noisy measurements.} 
		\label{simulation results6} 
	\end{figure}

	In this section, numerical simulation results are provided to illustrate the performance of the proposed distributed hybrid feedback controllers \eqref{hybridcontroller} and \eqref{velocityfreecontroller}, referred to as “Basic Hybrid" and “Velocity-Free Hybrid", respectively. For comparison purposes, we include the continuous feedback controller, referred to as “Non-Hybrid":
	\begin{align} 
		u_{i} &    =-k_X  
		\sum\nolimits_{\kappa \in \mathcal{M}_i} \bar{B}_{i\kappa} \bar{\psi}((I_4-\bar{X}_\kappa^{-1})\mathbb{A})
		-K_{\xi}\xi_i \nonumber \\
		&  ~~~~-K_e\sum\nolimits_{j \in \mathcal{N}_i}(\xi_i-\xi_j)  \nonumber 
	\end{align}
	which is modified from the distributed hybrid controller \eqref{hybridcontroller} by taking $\theta_k \equiv 0$ for all $k\in \mathcal{M}_i^+$.
	\begin{figure}[!ht]
		\centering	\includegraphics[width=0.48\textwidth]{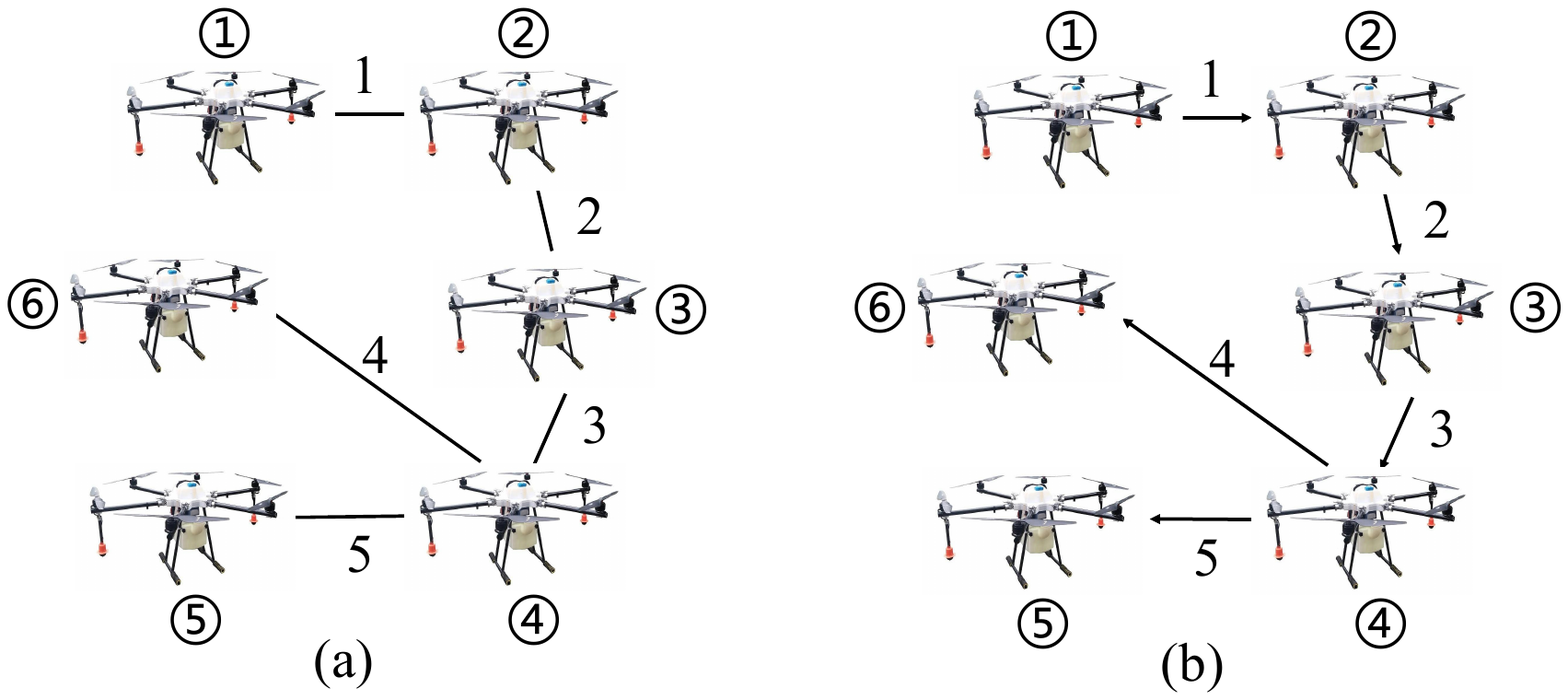}
		\vspace{-0.6in}
		\caption{The topology graph $\mathcal{G}$: (a) undirected; (b) undirected graph with an arbitrary orientation.} 
		\label{fig:graph} 
		\centering
		\includegraphics[width=0.48\textwidth]{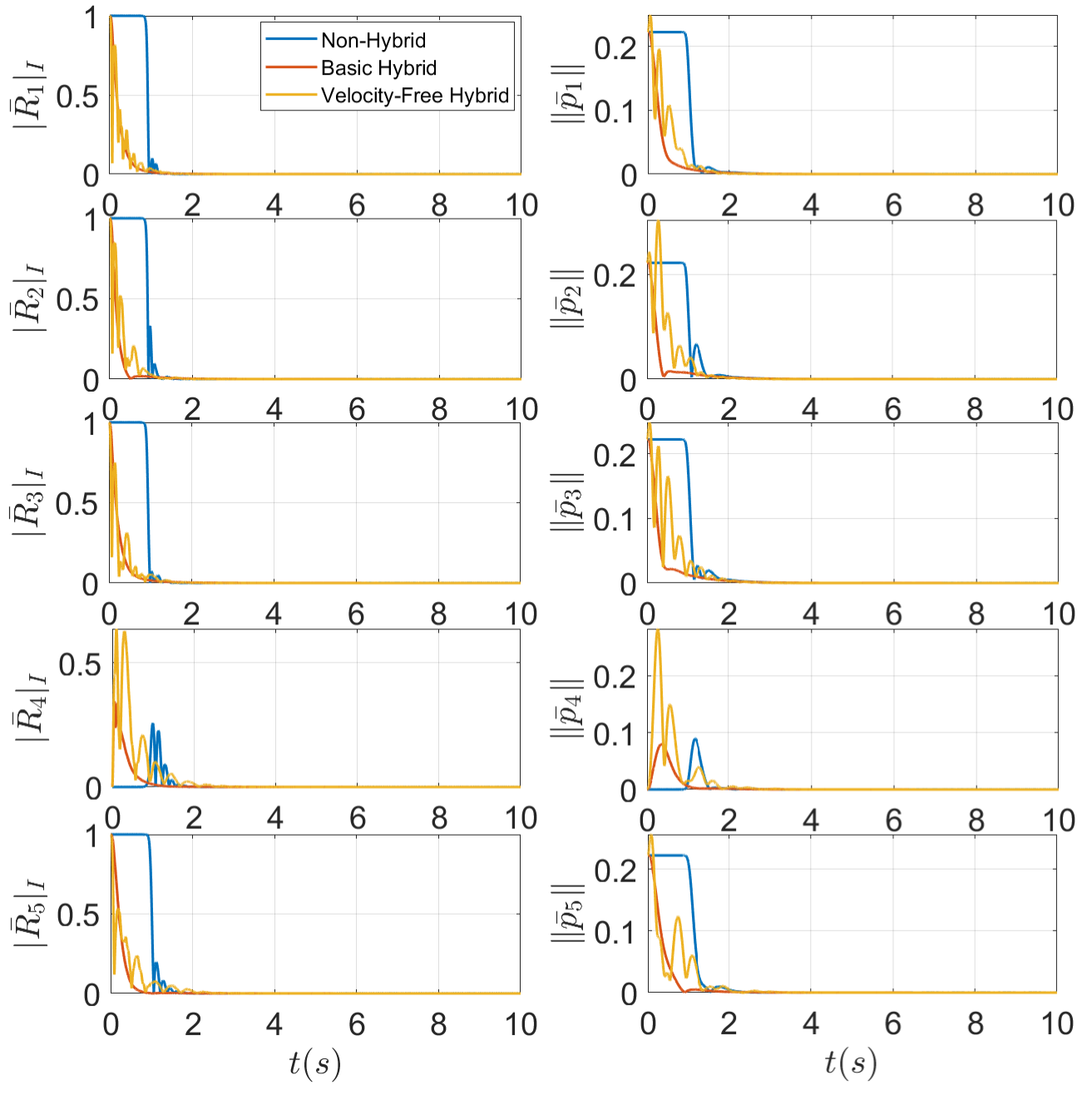}
		\vspace{-0.3in}
		\caption{Time evolution of the relative attitude and position associated with each edge.} 
		\label{simulation results1} 
	\end{figure}
	We consider six-rigid body systems with the undirected communication topology shown in  Fig.~\ref{fig:graph} (a). In Fig.~\ref{fig:graph} (b), we assign an arbitrary orientation to the graph $\mathcal{G}$ for the implementation of the dynamics of $\theta_k$ at the positive node of edge $k \in \mathcal{M}$. The neighbour sets are given as
	$\mathcal{N}_1=\{2\}, \mathcal{N}_2=\{1,3\}, \mathcal{N}_3=\{2,4\}, \mathcal{N}_4=\{3,5,6\}, \mathcal{N}_5=\{4\}$ and $\mathcal{N}_6=\{4\}$. 
	The inertia and mass parameters of the heterogeneous rigid body systems are taken as $J_i=(\diag([0.032,0.030,0.061])+0.020i I_3)\kg\cdot \m^2, m_i=(1.8+0.2i)\kg, i \in \mathcal{V}$. In our first simulation, we consider the matrix $\mathbb{A}$ in \eqref{AA_def} for the proposed potential function $U$ in  \eqref{U_def} as $A = \blkdiag(3,4,6)$, $b=[0~1~0]\T, d=9$, and choose the set of parameters $\mathcal{P}_{\mathbb{A}}$ in \eqref{eqn:P_A} as 
	$\Theta=\{ -0.9 \pi, +0.9 \pi\}$,   $u_{c1}=[0.346,0.566,0.748]^{\top}$,  $u_{c2} =-u_{c1}^{\times}bd^{-1} = [0.083,0,-0.038]^{\top}$,  
	$\gamma=\frac{10}{\pi^2}<\frac{4\Delta_W^*}{\pi^2}=\frac{560}{53\pi^2}$, and $\delta =0.21<\big(\frac{4\Delta_W^*}{\pi^2}-\gamma\big)\frac{\theta_{\max}^2}{2}= \frac{243}{1060} \approx 0.229$ with $\Delta_W^*={4\lambda_1^W\lambda_2^W\lambda_3^W}/{\sum_{\imath=1}^3 \sum_{k\neq \imath} \lambda_{\imath}^W \lambda_k^W}=2.642$.
	In addition, we consider the following initial conditions: 
	$ \theta_k(0)=\tilde{\theta}_i(0)=0$, $\xi_i(0)=\mathbf{0}$, $R_i(0)=\mathcal R_a((-1)^i(\pi/2+\epsilon),v), \hat{R}_i(0)=\mathcal R_a((-1)^{i+1}\pi/2-\epsilon,v)$ for $k\in \mathcal{M}$ and $i\in \mathcal{V}$
	with $v=[ 0, 0, 1]^{\top}$ and $\epsilon=10^{-12}$, $p_1(0)=\hat{p}_1(0)=[1,2,3]^{\top}$, $p_6(0)=p_4(0)$ and $\hat{p}_6(0)=\hat{p}_4(0)$, 
	and the other initial relative states $(\bar{X}_k(0), \theta_k(0)), k\in\{1,2,3,5\}$ are chosen close to one of the undesired equilibria.
	The gain paramter for all the controllers are set to $ k_{\theta} = k_{\tilde{\theta}} = 50, k_X = 20, k_{\tilde{X}}=150, K_{\xi}=\blkdiag(I_3,20I_3),
	K_e=\blkdiag(I_3,2I_3)$, and
	$K_{\hat{X}} = \blkdiag(300I_3, 15I_3)$.
	The second simulation considers the presence of measurement noise. For each $i \in \mathcal{V}$, the noisy measurements of pose and group velocities are modeled as $X_{iy}=X_i \exp(n_{iX}^\wedge), 
	\xi_{iy}=\xi_i+n_{i \xi}$, with $n_{iX}$ and
	$n_{i\xi}$ denoting independently zero-mean Gaussian distributions given by $n_{iX} \sim \mathcal{N}(0,\blkdiag(4\times10^{-4}I_3,2.5\times10^{-5}I_3)),n_{i\xi}\sim \mathcal{N}(0,\blkdiag
	(2.5\times10^{-3}I_3,2.5\times10^{-5}I_3))$.
	\begin{figure}[!ht]
		\centering
		\includegraphics[width=0.48\textwidth]{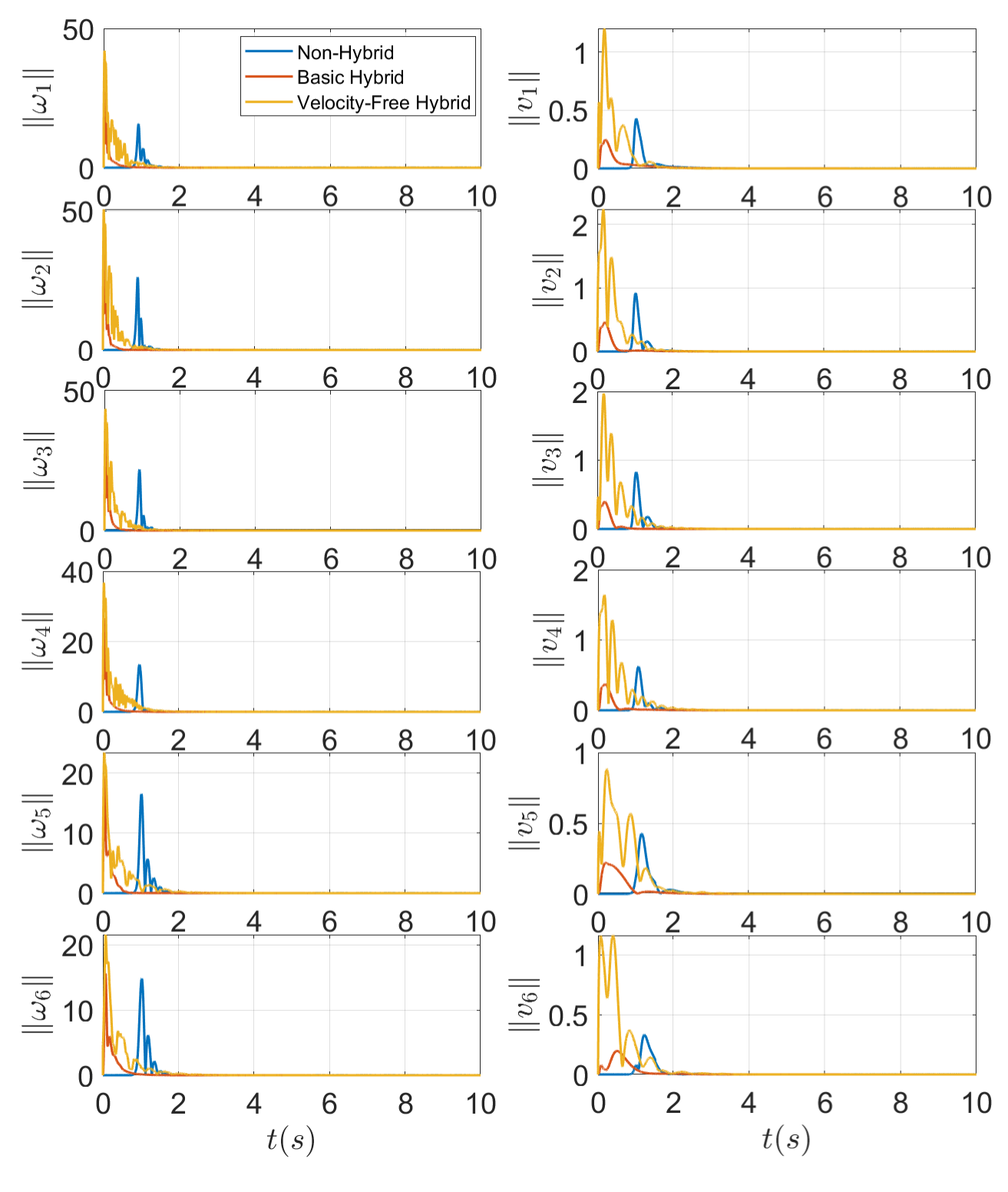}
		\label{simulation results2} 
		\vspace{-0.3in}
		\caption{Time evolution of the angular and linear velocity associated with each rigid body.} 
		\centering
		\includegraphics[width=0.48\textwidth]{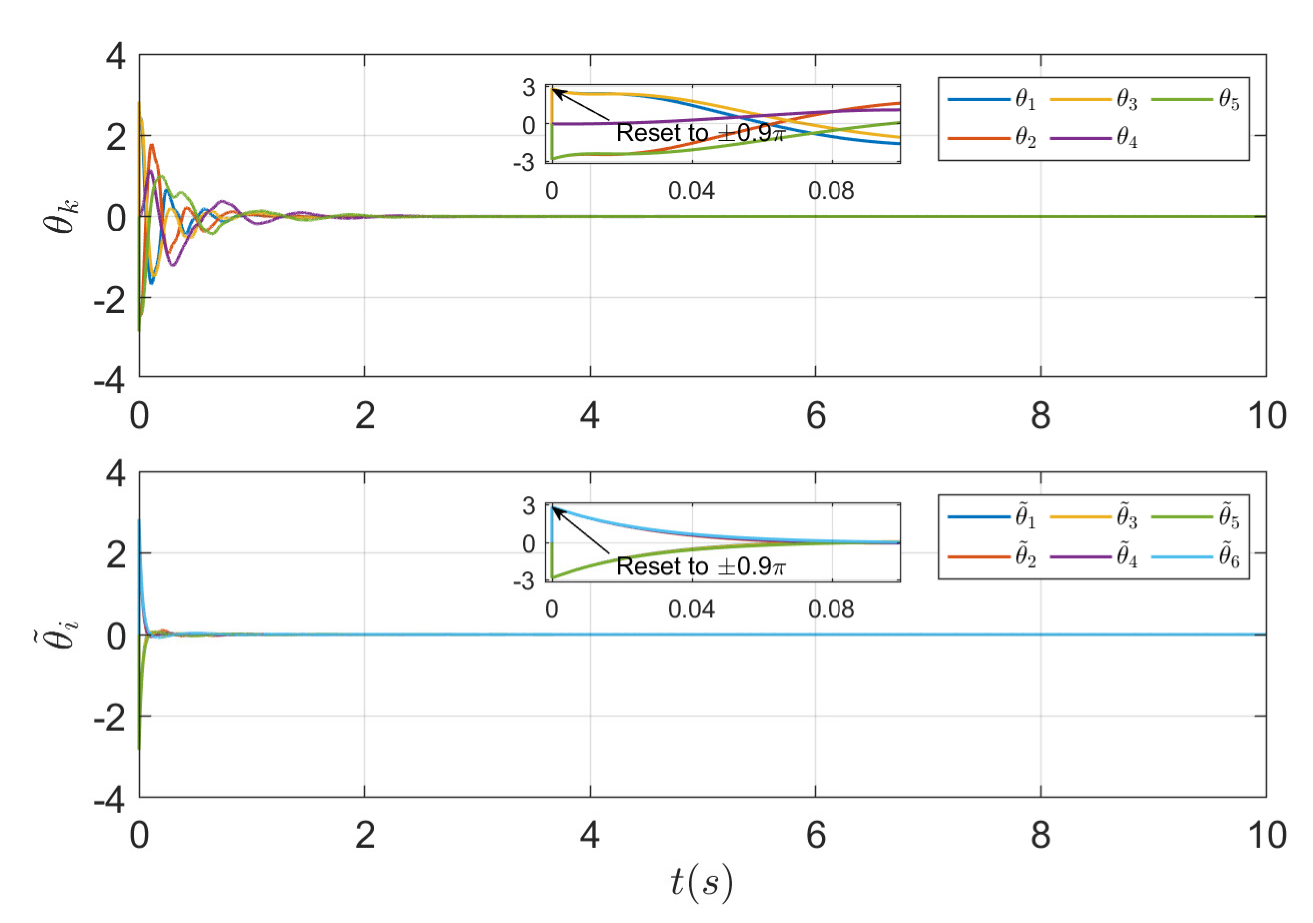}
		\vspace{-0.3in}
		\caption{Time evolution of the switching variables $\theta_k$ and $\tilde{\theta}_i$.} 
		\label{simulation results3} 
	\end{figure}
	From Figs.~\ref{simulation results1}-\ref{simulation results3}, one can see that the error states $(|\bar{R}_k|_I, \bar{p}_k)$ and the group velocity $(\omega_i, v_i)$ 
	converge to zero as $t\rightarrow \infty$. the auxiliary variables $\theta_k, \forall k \in \{1,2,3,5\}$, as well as auxiliary variables $\tilde{\theta}_i, i \in \mathcal{V}$ (for the velocity-free hybrid controller) jump from $0$ to $\pm0.9\pi$  and then converge to zero as $t\rightarrow \infty$. From Figs.~\ref{simulation results4}-\ref{simulation results6}, we can see that the hybrid controllers improve the convergence rate compared with the nonhybrid controller, when the error states are initialized around the undesired equilibrium point in the noise-free case. Moreover, these three controllers in the presence of measurement noise exhibit similar performance. Note that the velocity-free hybrid controller is more sensitive to noise mainly due to the large gains $K_{\hat{X}}$ involved in the auxiliary system \eqref{dyncheckX} and $k_{\tilde{X}}$ in controller \eqref{velocityfreecontroller} to overcome the lack of group velocity measurements.
	
	\section{Conclusion}
	We proposed two types of distributed hybrid pose synchronization schemes on $\SE(3)$ with/without group velocity measurements, with GAS guarantees, over undirected, connected, and acyclic graphs. A new potential function on $(\SE(3) \times \mathbb{R})^{M}$, involving a set of switching variables $\theta_k, k\in \mathcal{M}$ with flow and jump dynamics, was constructed to help with the design of the proposed hybrid control schemes. The first distributed hybrid synchronization scheme relies on the relative pose and the group velocity measurements. 
	However, such a control scheme requires group velocity measurements and cannot be directly applied to the scenario where the velocity measurements are not available. To overcome this problem, we proposed a distributed hybrid pose synchronization scheme without velocity measurements, relying on a hybrid auxiliary system for each rigid body to generate the necessary damping that compensates for the lack of group velocity information. 
	A natural extension of this work is to extend the proposed global pose synchronization strategy to synchronization/formation control for underactuated rigid body systems, over general communication graph topologies, including directed graphs. 

	\section*{Appendix}
	\appendix
	
	\section{Proof of Lemma \ref{lem: prop_barB} }\label{sec:lem_HE} 
	From the definitions of $\bar{\xi}_k=\xi_i-\Ad_{\bar{X}_k}^{-1}\xi_j$  and $\bar{B}(t)$ in \eqref{Bt_def}, one can obtain  $\bar{\xi}_k= \sum_{i=1}^N \bar{B}_{ik}^{\top} \xi_i$ for all $k \in \mathcal{M}_i$ with $i \in \mathcal{V}, j\in \mathcal{N}_i$. Then, it follows that $\bar{\xi}=\bar{B}(t)^{\top}\xi$, which implies that condition $1)$ in Lemma \ref{lem: prop_barB}  holds. Next, we present the proof of condition $2)$ in Lemma \ref{lem: prop_barB}. Inspired by \cite{boughellaba2026global}, we will prove Lemma \ref{lem: prop_barB} by contradiction.  Note that the graph is connected and acyclic, which implies that the graph contains a tree and the number of edges is $M=N-1$.  
	Then, one has rank$(\bar{B}(t)) \leq 6N-6$ due to $\bar{B}(t) \in \mathbb{R}^{6N \times 6M}$. Moreover, let $\xi=[\xi_1^{\top},\dots,\xi_N^{\top}]^{\top} \in \mathbb{R}^{6N}$ be a vector that belongs to the null space of $\bar{B}^{\top}(t), \ie,  \bar{B}^{\top}(t)\xi=\mathbf{0}$. According to \eqref{Bt_def}, one has $\xi_j=\Ad_{\bar{X}_k}\xi_i$ owing to the invertibility of $\Ad_{\bar{X}_k}$, for all $(i,j)\in \mathcal{E}$ and $k\in \mathcal{M}_i^+ \cap \mathcal{M}_j^-$. 
	For each vertex \(j\neq1\), let \(i_0=1,i_1,\ldots,i_{s_j}=j\) denote the unique path from vertex \(1\) to vertex \(j\), and let \(k_r\) denote the edge connecting \(i_{r-1}\) and \(i_r\). Define $\mathcal{Q}_1=I_6$ and $\mathcal{Q}_j
	=
	\Ad_{\bar X_{k_{s_j}}}^{\varepsilon_{j,s_j}}
	\cdots
	\Ad_{\bar X_{k_1}}^{\varepsilon_{j,1}}$, where
	\begin{align*}
		\varepsilon_{j,r}=
		\begin{cases}
			1,
			&k_r\in\mathcal M_{i_{r-1}}^+\cap\mathcal M_{i_r}^-,\\
			-1,
			&k_r\in\mathcal M_{i_{r-1}}^-\cap\mathcal M_{i_r}^+.
		\end{cases}
	\end{align*}
	Due to the connectivity of the graph $\mathcal{G}$,
	it yields $\xi_j=\mathcal{Q}_j (\Ad_{\bar{X}_1}, \dots, \Ad_{\bar{X}_M})\xi_1$, for each $j\in \mathcal{V}$. Define $\mathcal{Q}: =[\mathcal{Q}_1^{\top}, \dots, \mathcal{Q}_N^{\top}]^{\top} \in \mathbb{R}^{6N \times 6}$, one can show that $\xi=\mathcal{Q}(\Ad_{\bar{X}_1}, \dots, \Ad_{\bar{X}_M})\xi_1$. 
	The injectivity of $\mathcal{Q}$, which follows from the invertibility of each $\mathcal{Q}_j$, implies that the solution space of $\bar{B}(t)^{\top} \xi = \mathbf{0}$ is isomorphic to $\xi_1 \in \mathbb{R}^6$.  Therefore, it is clear that $\dim \nulls (\bar{B}^{\top}(t))=6$. Together with rank-nullity theory, one can obtain $\rank(\bar{B}(t))=\rank(\bar{B}(t)^{\top})=6N-6$, which means that the matrix $\bar{B}(t)$ is full of column rank. It follows that the only solution to $\bar{B}(t)y=\mathbf{0}$ with $y\in \mathbb{R}^{6M}$ is $y=\mathbf{0}$.  This completes the proof.

	\section{Proof of Lemma \ref{lem:parameters}} \label{sec:lem_parameters} 
	The proof of this lemma is inspired by \cite[Proposition 2]{wang2022hybrid}. 
	From  \eqref{def:Upsilon_V} and \eqref{U_cri}, the set of undesired critical points is given by $ \Upsilon_U \setminus \{\mathcal{A}_0\} = \{(\bar{X}_k,\theta_k) \in \SE(3) \times \mathbb{R}: \bar{X}_k=\mathcal{T}(\bar{R}_k, \bar{p}_k), \bar{R}_k=\mathcal{R}_a(\pi, v), \bar{p}_k=(I_3-R_a(\pi, v))bd^{-1}, v\in \mathcal{E}_v(W), \theta_k=0\}$. This implies that, for any $(\bar{X}_k, \theta_k) \in \Upsilon_U \setminus \{\mathcal{A}_0\}$, one has $\theta_k = 0$, $\bar{X}_k=\mathcal{T}(\bar{R}_k, \bar{p}_k)$ with $ \bar{R}_k=\mathcal{R}_a(\pi, v), \bar{p}_k =(I_3-R_a(\pi, v))bd^{-1}$ and $v \in \mathcal{E}_v(W)$. Then, from \eqref{U_explicit}, for any $(\bar{X}_k, \theta_k) \in \Upsilon_U \setminus \{\mathcal{A}_0\}$ with $\bar{X}_k=\mathcal{T}(\bar{R}_k, \bar{p}_k)$ and $\theta_k' \in \Theta \subset \mathbb{R}$, one obtains
	\begin{align}\label{U(pi,v,),0}
		U(\bar{X}_k, \theta_k)
		&  =\tr(W(I_3-\bar{R}_k\mathcal{R}_a(\theta_k, u_{c1})))+\frac{\gamma}{2}\theta_k^2 \nonumber\\
		&  ~~~~+\frac{d}{2}\|\bar{p}_k - (I_3 - \bar{R}_k)bd^{-1}\|^2 \nonumber \\
		&=\tr(W(I_3-\mathcal{R}_a(\pi,v))) 
	\end{align}
	and
	\begin{align}\label{U(theta,v,),theta}
		U(\bar{X}_k,\theta_k') 
		&  =\tr(W(I_3-\bar{R}_k\mathcal{R}_a(\theta_k', u_{c1})))+\frac{\gamma}{2}\theta_k'^2 \nonumber\\
		&  ~~~~+\frac{d}{2}\|\bar{p}_k - (I_3 - \bar{R}_k)bd^{-1}\|^2 \nonumber \\
		&  =\tr(W(I_3-\mathcal{R}_a(\pi,v)\mathcal{R}_a(\theta_k',u_{c1}))) +\frac{\gamma}{2}\theta_k'^2  \nonumber\\
		&=\tr(W(I_3-\mathcal{R}_a(\pi,v))) \nonumber \\
		&  ~~~~ +\tr(W\mathcal{R}_a(\pi,v)(I_3-\mathcal{R}_a(\theta_k',u_{c1})))+\frac{\gamma}{2}\theta_k'^2  \nonumber\\
		&  =U(\bar{X}_k,\theta_k)-2\sin^2(\frac{\theta_k'}{2})\Delta(v,u_{c1}) +\frac{\gamma}{2}\theta_k'^2  
	\end{align}
	where we have made use of the facts: $\mathcal{R}_a(\theta_k',u_{c1})=I_3+\sin (\theta_k') u_{c1}^{\times}+(1-\cos(\theta_k'))(u_{c1}^{\times})^2$ and
	$\tr(W\mathcal{R}_a(\pi,v)(I_3-\mathcal{R}_a(\theta_k',u_{c1}))
	=-2\sin^2(\frac{\theta_k'}{2})\Delta(v,u_{c1})$ with $\Delta(v,u_{c1})=u_{c1}^{\top}(\tr(W)I_3-W-2v^{\top}Wv(I_3-vv^{\top}))u_{c1}$ for all $\theta_k' \in \mathbb{R}, v\in \mathcal{E}_v(W), u_{c1}\in \mathbb{S}^2$. Given the choice of $u_{c1} \in \mathbb{S}^2$ as per the set $\mathcal{P}_{\mathbb{A}}$ in \eqref{eqn:P_A}, it  follows from \cite[Proposition 2]{berkane2017construction} that $\Delta^*=\min_{v\in \mathcal{E}_v(W)}\Delta(v,u_{c1})>0$.

	Therefore, from \eqref{U(pi,v,),0} and \eqref{U(theta,v,),theta}, for any $(\bar{X}_k, \theta_k) \in \Upsilon_U \setminus \{\mathcal{A}_0\}$ and $\Theta = \{|\theta_{\jmath}| \in (0,\pi], \jmath=1,\dots,m\} \subset \mathbb{R}$, the following inequality holds
	\begin{align}
		\mu_U(\bar{X}_k, \theta_k)&  =U(\bar{X}_k,\theta_k)-\min_{\theta_k' \in \Theta} U(\bar{X}_k,\theta_k')\nonumber\\
		&   =\max_{\theta_k' \in \Theta}\left(2\sin^2\big(\frac{\theta_k'}{2})\Delta(v,u_{c1}\big)-\frac{\gamma}{2}\theta_k'^{2}\right)\nonumber\\
		& \geq \max_{\theta_k' \in \Theta}\left(2\sin^2\big(\frac{\theta_k'}{2}\big)\Delta^*-\frac{\gamma}{2}\theta_k'^{2}\right)\nonumber\\
		& \geq 2\sin^2\big(\frac{\theta_{\max}}{2}\big)\Delta^*-\frac{\gamma}{2}\theta_{\max}^2 \nonumber \\
		& \geq \left(\frac{4\Delta^*}{\pi^2}-\gamma \right)\frac{\theta_{\max}^2}{2}>\delta
	\end{align}
	where we made use of the facts $\frac{4\Delta^*}{\pi^2}-\gamma>0$, $\delta<(\frac{4\Delta^*}{\pi^2}-\gamma)\frac{\theta_{\max}^2}{2}$ and $\theta_{\max}=\sup_{\theta_k' \in \Theta}|\theta_k'| \leq \pi$ given by \eqref{eqn:P_A}, and $2\sin^2(\frac{\theta_k'}{2})\Delta^*-\frac{\gamma}{2}\theta_k'^{2}> 0$, {$|\sin(\frac{\theta'_k}{2})| \geq \frac{|\theta'_k|}{\pi}$ for all $|\theta_k'|\in (0,\pi]$ (\ie, $\theta_k' \in \Theta$). This completes the proof.

		\section{Proof of Lemma~\ref{lem:hybrid basic condition}}\label{sec:lem_hybrid_basic_condition}
		Regarding each pose matrix as a vector by stacking its entries, the state space $\mathcal S_c=(\mathrm{SE}(3)\times\mathbb R)^M\times\mathbb R^{6N}$ is a closed subset of $\mathbb{R}^{17M+6N}$.
		Since $\Theta$ is finite and $U$ is continuous, $\mu_U$ is continuous (the composition of continuous functions). Therefore, the flow set $\mathcal{F}_c$ and the jump set $\mathcal{J}_c$ in \eqref{hybsys} are closed subsets of $\mathcal{S}_c=(\SE(3) \times \mathbb{R})^M \times \mathbb{R}^{6N} \subseteq \mathbb{R}^{17M+6N}$ and satisfy $\mathcal{F}_c \cup \mathcal{J}_c=\mathcal{S}_c$. This indicates that condition $A1$ holds. Since $f$ and $u_i$ are single-valued continuous functions on $\mathcal{F}_c$, the properties of outer semicontinuity, local boundedness, and convexity hold for the flow map $F$ in \eqref{hybsys}. This indicates that condition $A2$ holds.\\
		To verify condition $A3$, we first show the outer semicontinuity of $g$ in \eqref{eqn: gtheta}. 
		For each edge $k \in \mathcal{M}$, consider a sequence $\{(\bar X_k^r,\theta_k^r)\}_{r \in \mathbb{N}} \subseteq \SE(3) \times \mathbb{R}$ converging to $(\bar X_k,\theta_k)$, and a sequence $\{\eta_k^r\}_{r \in \mathbb{N}}$ satisfying $U(\bar{X}_k^r, \eta_k^r)= \min_{\theta'_k \in \Theta} U(\bar{X}_k^r, \theta_k')$ ($\ie, \eta_k^r\in g(\bar{X}_k^r,\theta_k^r)= \arg\min_{\theta_k' \in \Theta}U(\bar{X}_k^r, \theta'_k)$) for each $r \in \mathbb{N}$. Since  $\Theta$ is compact (finite non-empty subset of~$\mathbb{R})$ by assumption, there exists a subsequence $\{\eta_k^{r(s)}\}_{s \in \mathbb{N}}$ that converges to some $\eta_k \in \Theta$. 
		For every $\theta'_k \in \Theta$, one has $U(\bar{X}_k^{r(s)}, \eta_k^{r(s)}) \leq U(\bar{X}_k^{r(s)}, \theta'_k)$ by the definition of $g$ in \eqref{eqn: gtheta}. Together with the continuity of $U$, it follows that $U(\bar{X}_k, \eta_k) \leq U(\bar{X}_k, \theta'_k)$ for every $\theta'_k \in \Theta$. Hence, $\eta_k \in g(\bar{X}_k, \theta_k)$, which implies that  the set-valued function $g$ is outer semicontinuous.\\
		Next, consider a sequence $\{x_c^r\} \subseteq \mathcal{J}_c$ converging to $x_c \in \mathcal{J}_c$, and a sequence $y_c^r \in G(x_c^r)$ converging to $y_c$ for each $r \in \mathbb{N}$, where $y_c=(\bar{X}_1, \eta_1, \dots, 
		\bar{X}_M, \eta_M, \xi_1, \dots, \xi_N)$ and $y_c^r=(\bar{X}_1^r, \eta_1^r, \dots, 
		\bar{X}_M^r, \eta_M^r, \xi_1^r, \dots, \xi_N^r)$. For each  $k \in \mathcal{M}$, there exists a subsequence  such that $\eta_k^{r(s)} \in g(\bar{X}_k^{r(s)}, \theta_k^{r(s)})$ and $\mu_U(\bar{X}_k^{r(s)}, \theta_k^{r(s)}) \geq \delta$ or $\eta_k^{r(s)} = \theta_k^{r(s)}$ and $\mu_U(\bar{X}_k^{r(s)}, \theta_k^{r(s)}) \leq \delta$, with $g$ and $\mu_U$ given by \eqref{eqn: gtheta} and \eqref{mu_U_def}, respectively. By the outer semicontinuity of $g$ and the continuity of $\mu_U$, one obtains $\eta_k \in g_{\theta}(\bar{X}_k, \theta_k)$ for every $k \in \mathcal{M}$. Hence, one has $y_c \in G(x_c)$. Consequently $G$ is outer semicontinuous relative to $\mathcal{J}_c$.\\
		Furthermore, since the set $\Theta$ is finite and all states except the switching variable remain unchanged, one has $G$ is locally bounded. Finally, for every $x_c \in \mathcal{J}_c$, there exists at least one $k$ satisfying $\mu_U(\bar{X}_k, \theta_k) \geq \delta$, which implies $G(x_c) \neq \emptyset$ and thus $\mathcal{J}_c \subseteq \dom G$. This indicates that the condition $A3$ holds, which completes the proof.
		
		\section{Proof of Theorem~\ref{T1}}\label{sec:pfT1}
		Consider the following Lyapunov function candidate:
		\begin{align}\label{V_def}
			\bar{V}(x_c)=k_X\bar{U}(x)+\sum\nolimits_{i=1}^N \xi_i^{\top} \mathbb{I}_i \xi_i.
		\end{align}
		From \eqref{eqn:errdynamics}, \eqref{U_def}, \eqref{barU_def}, \eqref{eqn: gtheta} and Lemma \ref{lem:Uproperty}, the time-derivative of $U(\bar{X}_k, \theta_k)$ is given by
		\begin{align*} 
			&\dot{U}(\bar{X}_k, \theta_k) \nonumber \\
			&\quad =\langle \nabla_{\bar{X}_k}U(\bar{X}_k, \theta_k), \bar{X}_k\bar{\xi}_k^{\wedge} \rangle_{\bar{X}_k} +\langle\langle \nabla_{\theta_k}U(\bar{X}_k, \theta_k), \dot{\theta}_k \rangle\rangle  \nonumber \\
			&\quad=\langle\langle \bar{X}_k^{-1}\nabla_{\bar{X}_k}U(\bar{X}_k, \theta_k), \bar{\xi}_k^{\wedge} \rangle\rangle+\nabla_{\theta_k}U(\bar{X}_k, \theta_k)\dot{\theta}_k  \nonumber  \\
			&\quad=2\bar{\xi}_k^{\top} \bar{\psi}_{\nabla}(\bar{X}_k, \theta_k)-k_{\theta}|\nabla_{\theta_k}U(\bar{X}_k, \theta_k)|^2
		\end{align*}
		where we made us of the fact $\langle\langle \mathbb{A}, x^{\wedge} \rangle\rangle=2x^{\top}\bar{\psi}(\mathbb{A})$, it follows from \eqref{barU_def} that
		\begin{align}\label{eqn:dot_bar_U}
			\dot{\bar{U}}(x)
			& =\sum_{k=1}^M (2\bar{\xi}_k^{\top} \bar{\psi}_{\nabla}(\bar{X}_k, \theta_k)-k_{\theta}|\nabla_{\theta_k}U(\bar{X}_k, \theta_k)|^2) \nonumber \\
			&=2\bar{\xi}^{\top}\Psi^{\bar{X}}_{\nabla}(x)-k_{\theta}\|\Psi_{\nabla}^{\theta}(x)\|^2 \nonumber \\ 
			&=2\xi^{\top}\bar{B}\Psi^{\bar{X}}_{\nabla}(x)-k_{\theta}\|\Psi_{\nabla}^{\theta}(x)\|^2
		\end{align}
		where $\Psi^{\bar{X}}_{\nabla}(x):=[\bar{\psi}_{\nabla}(\bar{X}_1, \theta_1)^{\top},\dots,
		\bar{\psi}_{\nabla}(\bar{X}_M, \theta_M)^{\top}]^{\top} 
		\in \mathbb{R}^{6M}$ and $\Psi_{\nabla}^{\theta}(x):=[\nabla_{\theta_1}U(\bar{X}_1, \theta_1),\dots,
		\nabla_{\theta_M}U(\bar{X}_M, \theta_M)]^{\top}\\ \in \mathbb{R}^M$ and  we made use of the facts: the gradients of $\bar{U}$ in \eqref{barU_def} and $\bar{\xi}=\bar{B}^{\top}\xi$ as per Lemma \ref{lem: prop_barB}. 
		
		From \eqref{eqn:errdynamics} and \eqref{hybridcontroller}, one can show that
		\begin{align}\label{eqn:sum_dot_xi_i}
			\sum_{i=1}^N \xi_i^{\top}\mathbb{I}_i \dot{\xi}_i
			& = -k_X\sum_{i=1}^N \xi_i^{\top} \sum_{\kappa \in \mathcal{M}_i}\bar{B}_{i\kappa} \bar{\psi}_{\nabla}(\bar{X}_{\kappa}, \theta_{\kappa})\nonumber \\
			& ~~~~ -\sum_{i=1}^N \xi_i^{\top} \big(K_{\xi}\xi_i + K_e \sum_{j \in \mathcal{N}_i}(\xi_i-\xi_j) \big)  \nonumber \\
			& =-k_X\sum_{i=1}^N \xi_i^{\top} \sum_{\kappa =1}^M\bar{B}_{i\kappa} \bar{\psi}_{\nabla}(\bar{X}_{\kappa}, \theta_{\kappa})\nonumber \\  
			& ~~~~ -\sum_{i=1}^N \xi_i^{\top} \big(K_{\xi}\xi_i + K_e \sum_{j \in \mathcal{N}_i}(\xi_i-\xi_j) \big)  \nonumber \\
			& = -k_X\xi^{\top}\bar{B}(t)\Psi^{\bar{X}}_{\nabla}(x)- \xi^{\top}(I_N \otimes K_{\xi}) \xi \nonumber \\  
			& ~~~~ -\xi^{\top}(\mathcal{L}  \otimes K_e) \xi
		\end{align}
		where $\mathcal{L}:= BB^{\top} \in \mathbb{R}^{N\times N}$ is known as the Laplacian matrix with $B$ defined in Section \ref{sec:graph_theory}, and we made use of the facts: $\ad_{\xi_i}\xi_i=\mathbf{0}$, $\mathcal{M}=\{1,\dots, M\}$, $\sum_{\kappa \in \mathcal{M}_i}\bar{B}_{i\kappa} \bar{\psi}_{\nabla}(\bar{X}_{\kappa}, \theta_{\kappa}) = \sum_{\kappa \in \mathcal{M}}\bar{B}_{i\kappa} \bar{\psi}_{\nabla}(\bar{X}_{\kappa}, \theta_{\kappa})=\sum_{\kappa =1}^M \bar{B}_{i\kappa} \bar{\psi}_{\nabla}(\bar{X}_{\kappa}, \theta_{\kappa})$ from the definition of $\bar{B}$ in \eqref{Bt_def}, and $\sum_{i=1}^N \xi_i^{\top} K_e \sum_{j \in \mathcal{N}_i}(\xi_i-\xi_j)=\xi^{\top}(\mathcal{L} \otimes K_e)\xi $ with $\otimes$ denoting the Kronecker product.
		Then, by virtue of \eqref{eqn:dot_bar_U} and \eqref{eqn:sum_dot_xi_i}, for all $x_c\in \mathcal{F}_c$ one obtains
		\begin{align}\label{eqn:re_dot_V}
			\dot{\bar{V}}(x_c) 
			&  = k_X\dot{\bar{U}}(x)+2\sum_{i=1}^N \xi_i^{\top} \mathbb{I}_i \dot{\xi}_i \nonumber \\
			& = -k_Xk_{\theta}\|\Psi_{\nabla}^{\theta}(x)\|^2- 2\xi^{\top}(I_N \otimes K_{\xi}) \xi \nonumber \\  
			& ~~~~ -2\xi^{\top}(\mathcal{L}  \otimes K_e) \xi \leq 0.
		\end{align}
		This implies that $\bar{V}$ is non-increasing along the flows of  \eqref{hybsys}.

		On the other hand, for each jump $x_c \in \mathcal{J}_c$, it follows from the definition of $\mathcal{J}_c$ that there exists a nonempty subset $\mathcal{M}' \subseteq \mathcal{M}$ such that at least one edge $k \in \mathcal{M}$ undergoes a jump satisfying $\mu_U(\bar{X}_k, \theta_k) \geq \delta$ with $\mu_U$ given by \eqref{mu_U_def}. Together with \eqref{barU_def}, for all $x_c\in \mathcal{J}_c$ one has 
		\begin{align}\label{ineq:V_diff}
			&\bar{V}(x_c^+)-\bar{V}(x_c)=k_X\bar{U}(x^+)-k_X\bar{U}(x)  \nonumber\\
			& \qquad \quad  =-k_X \sum_{k\in \mathcal{M}'}(U(\bar{X}_k, \theta_k)-\min_{\theta_k' \in \Theta}U(\bar{X}_k, \theta'_k)) \nonumber \\
			& \qquad \quad =-k_X\sum_{k\in \mathcal{M}'} \mu_U(\bar{X}_k, \theta_k)  \nonumber \\ 
			& \qquad \quad \leq -k_X \sum_{k\in \mathcal{M}'} \delta\leq -k_X \delta.
		\end{align}
		This implies that $\bar{V}$ has a strict decrease over the jumps of \eqref{hybsys}. From \eqref{eqn:re_dot_V} and \eqref{ineq:V_diff}, one can conclude that $\bar{V}$ has compact sublevel sets. Since $\bar{V}$ is non-increasing along every solution, every solution remains in a compact sublevel set and is bounded. It follows from \cite[Theorem 23]{goebel2009hybrid} that $\bar{\mathcal{A}}$ is stable, and every maximal solution to the hybrid closed-loop system \eqref{hybsys} is bounded. One can obtain from \eqref{eqn:re_dot_V} and \eqref{ineq:V_diff} that $\bar{V}(x_c(t,j)) \leq \bar{V}(x_c(t_j,j))  \leq \bar{V}(x_c(t_j,j-1))-k_X\delta$ for all $(t,j),(t_j,j),(t_j,j-1) \in \dom x_c$ with $(t,j) \succeq (t_j,j) \succeq (t_j,j-1)$. 
		Therefore, it is clear that $0 \leq \bar{V}(x_c(t,j)) \leq \bar{V}(x_c(0,0))-j k_X\delta$ for all $(t,j)\in \dom x_c$, which leads to  $j \leq \lceil \bar{V}(x_c(0,0))/k_X\delta \rceil$ with $\lceil \cdot \rceil$ denoting the ceiling function. 
		Thus, one concludes that the number of jumps is finite and depends on the initial conditions. 
		
		Now, we verify the viability condition. 
		For any
		$x_c\in\mathcal{F}_c\setminus\mathcal{J}_c$, one has
		$\mu_U(\bar{X}_k,\theta_k)<\delta$ for all
		$k\in\mathcal{M}$ from \eqref{set_Fk_J_k} and \eqref{hybsys}. Since $\mu_U$ is continuous  and
		$\mathcal{M}$ is finite, there exists an open
		neighborhood $O$ of $x_c$ such that
		$O\cap\mathcal{S}_c\subseteq
		\mathcal{F}_c\setminus\mathcal{J}_c$.
		Together with the fact that $\mathcal{F}_c \subseteq \mathcal{S}_c$, one has $O\cap\mathcal{F}_c=O\cap\mathcal{S}_c$. Therefore,
		the sets $\mathcal{F}_c$ and $\mathcal{S}_c$ locally
		coincide, and hence
		$T_{\mathcal{F}_c}(x_c)=T_{\mathcal{S}_c}(x_c)$, where $T_{\mathcal{F}_c}(x_c)$ and $T_{\mathcal{S}_c}(x_c)$ denote the tangent cones to $\mathcal{F}_c$ and $\mathcal{S}_c$ at the point $x_c$. 
		For $\mathcal{S}_c$, one has $T_{\mathcal{S}_c}(x_c)
		=
		(T_{\bar{X}_1}\SE(3) \times\mathbb{R} ) \times\dots \times (T_{\bar{X}_M}\SE(3) \times\mathbb{R} )\times\mathbb{R}^{6N}$ with $T_{\bar{X}_k}\SE(3)$ denoting  the tangent space to $\SE(3)$ at $\bar{X}_k$. From the definition of $F(x_c)$ after \eqref{hybsys}, one has $F(x_c) \in T_{\mathcal{S}_c}(x_c)=T_{\mathcal{F}_c}(x_c)$ for any $x_c \in \mathcal{F}_c\setminus\mathcal{J}_c$.
		Note that every maximal solution to the hybrid system \eqref{hybsys} is bounded, $F(x_c) \in T_{\mathcal{F}_c}(x_c)$ for any $x_c \in \mathcal{F}_c \setminus \mathcal{J}_c$, 
		and $G(x_c) \subset \mathcal{F}_c \cup \mathcal{J}_c=\mathcal{S}_c$. Then, as per \cite[Theorem S3]{goebel2009hybrid}, it follows that every maximal solution to \eqref{hybsys} is complete, and therefore, it is complete and bounded,
		and thus precompact.
		
		Furthermore, applying the invariance principle for hybrid systems \citep{goebel2009hybrid}, it follows that every precompact solution to the hybrid closed-loop system \eqref{hybsys} must converge to the largest invariant set contained in $\varGamma:= \{x_c \in \mathcal{F}_c: \Psi_{\nabla}^{\theta}(x)=\mathbf{0}, \xi=\mathbf{0}\}$. For each $x_c\in \varGamma$, from $\xi \equiv \mathbf{0}$, it follows that $\dot{\xi}=\mathbf{0}$. This implies $\sum_{\kappa = 1}^M\bar{B}_{i\kappa} \bar{\psi}_{\nabla}(\bar{X}_{\kappa}, \theta_{\kappa})=\mathbf{0}$ (\ie, $\bar{B}(t)\Psi_{\nabla}^{\bar{X}}(x)=\mathbf{0}$) from \eqref{eqn:errdynamics} and \eqref{hybridcontroller}.
		Then, by virtue of Lemma \ref{lem: prop_barB}, it follows that $\Psi_{\nabla}^{\bar{X}}(x)=\mathbf{0}$. From $\Psi_{\nabla}^{\bar{X}}(x)=\mathbf{0}$ and $\Psi_{\nabla}^{\theta}(x)=\mathbf{0}$, one has $x \in \Upsilon_{\bar{U}}$ with $\Upsilon_{\bar{U}}$ denoting the set of critical point of $\bar{U}$. Consequently, any solution to the hybrid closed-loop system \eqref{hybsys} converges to the largest weakly invariant set contained in $\varGamma':=\{x_c \in \mathcal{S}_c: x \in \mathcal{F} \cap \Upsilon_{\bar{U}}, \xi=\mathbf{0}\}$.
		On the other hand, given $x \in \mathcal{A}$, one has, for all $k \in \mathcal{M}, \mu_U(\bar{X}_k, \theta_k)=-\min_{\theta_k' \in \Theta}U(\bar{X}_k, \theta_k') \leq 0$, which implies that $\mathcal{A} \subset \mathcal{F} \cap \Upsilon_{\bar{U}}$ and $\mathcal{F} \cap (\Upsilon_{\bar{U}} \setminus \mathcal{A})=\emptyset$. In addition, applying some set-theoretic arguments, one has $\mathcal{F} \cap \Upsilon_{\bar{U}} \subset (\mathcal{F} \cap (\Upsilon_{\bar{U}} \setminus \mathcal{A})) \cup (\mathcal{F} \cap \mathcal{A})=\emptyset \cup \mathcal{A}$. From $\mathcal{A} \subset \mathcal{F} \cap \Upsilon_{\bar{U}}$ and $\mathcal{F} \cap \Upsilon_{\bar{U}} \subset \mathcal{A}$, one has $\mathcal{F} \cap \Upsilon_{\bar{U}}=\mathcal{A}$. Thus, one can verify that $\varGamma'= \bar{\mathcal{A}}$.  
		Finally, from Lemma \ref{lem:hybrid basic condition}, one can conclude that the set $\bar{\mathcal{A}}$ is globally asymptotically stable. This completes the proof.

		\section{Proof of Theorem~\ref{T2}}\label{sec:pfT2}
		Consider the following Lyapunov function candidate:
		\begin{align}\label{vlofreeV_def}
			\tilde{V}(\tilde{x}_c)=k_X\bar{U}(x)+k_{\tilde{X}}\sum_{i=1}^N U(\tilde{X}_i, \tilde{\theta}_i)+\sum_{i=1}^N \xi_i^{\top} \mathbb{I}_i \xi_i.
		\end{align}
		Using the properties of the potential functions $\bar{U}$ and $U$, one can easily verify that $\tilde{V}$ is positive definite on $\tilde{\mathcal{S}}_c$ with respect to the closed set $\tilde{\mathcal{A}}$.
		From \eqref{U_def}, \eqref{eqn: gtheta}, \eqref{deqn:widetildeXdynamics} and Lemma \ref{lem:Uproperty}, one can obtain that 
		\begin{align}\label{lofreeUH_dev}
			&\dot{U}(\tilde{X}_i, \tilde{\theta}_i)\nonumber\\[2pt]
			&=\langle \nabla_{\tilde{X}_i} U(\tilde{X}_i, \tilde{\theta}_i), \tilde{X}_i\bar{\beta}_i^{\wedge} \rangle_{\tilde{X}_i} +\langle \nabla_{\tilde{\theta}_i} U(\tilde{X}_i, \tilde{\theta}_i), \dot{\tilde{\theta}}_i \rangle \nonumber\\
			&=\langle \langle \tilde{X}_i^{-1}\nabla_{\tilde{X}_i} U(\tilde{X}_i, \tilde{\theta}_i), \bar{\beta}_i^{\wedge} \rangle \rangle+\langle \nabla_{\tilde{\theta}_i} U(\tilde{X}_i, \tilde{\theta}_i), \dot{\tilde{\theta}}_i \rangle \nonumber\\
			&=2\bar{\beta}_i^{\top} \bar{\psi}_{\nabla}(\tilde{X}_i, \tilde{\theta}_i) +\dot{\tilde{\theta}}_i \nabla_{\tilde{\theta}_i} U(\tilde{X}_i, \tilde{\theta}_i) \nonumber\\
			&=2\bar{\beta}_i^{\top} \bar{\psi}_{\nabla}(\tilde{X}_i, \tilde{\theta}_i) -k_{{\theta}} |\nabla_{\tilde{\theta}_i} U(\tilde{X}_i, \tilde{\theta}_i)|^2 
		\end{align}
		where $ \bar{\beta}_i=\xi_i-\beta_i$, and we have used the fact $\langle\langle \mathbb{A}, x^{\wedge} \rangle\rangle=2x^{\top}\bar{\psi}(\mathbb{A})$. 
		It follows that
		\begin{align}\label{vlofreesumUH_dev}
			&\sum_{i=1}^N \dot{U}(\tilde{X}_i, \tilde{\theta}_i)  \nonumber\\
			&=2\sum_{i=1}^N \bar{\beta}_i^{\top} \bar{\psi}_{\nabla}(\tilde{X}_i, \tilde{\theta}_i)  -k_{{\theta}} \sum_{i=1}^N |\nabla_{\tilde{\theta}_i} U(\tilde{X}_i, \tilde{\theta}_i)|^2 \nonumber \\
			&=2 \bar{\beta}^{\top}\Psi_{\nabla}^{\tilde{X}}
			- k_{{\theta}} \|\Psi_{\nabla}^{\tilde{\theta}}\|^2
		\end{align}
		where 
		$\Psi^{\tilde{X}}_{\nabla}:=[ \bar{\psi}_{\nabla}(\tilde{X}_1, \tilde{\theta}_1)^{\top},\dots,
		\bar{\psi}_{\nabla}(\tilde{X}_N, \tilde{\theta}_N)^{\top}]^{\top} 
		\in \mathbb{R}^{6N}, \Psi_{\nabla}^{\tilde{\theta}}:=[\nabla_{\tilde{\theta}_1}U(\tilde{X}_1, \tilde{\theta}_1),
		\dots,
		\nabla_{\tilde{\theta}_N}U(\tilde{X}_N, \tilde{\theta}_N)]^{\top}  \in \mathbb{R}^N$, and $\bar{\beta}=[\bar{\beta}_1^{\top},  \dots, \bar{\beta}_N^{\top}]^{\top}  \in \mathbb{R}^{6N}$.
		
		In view of \eqref{eqn:kinematic} and \eqref{velocityfreecontroller}, one can show that
		\begin{align}\label{eqn:velocityfreesum_dot_xi_i}
			\sum_{i=1}^N \xi_i^{\top}\mathbb{I}_i \dot{\xi}_i&=\sum_{i=1}^N \xi_i^{\top}(\ad_{\xi_{i}}^{\top}\mathbb{I}_i\xi_i+u_{i}) \nonumber \\ 
			&=-k_X\sum_{i=1}^N \xi_i^{\top}\sum_{\kappa =1}^M\bar{B}_{i\kappa} \bar{\psi}_{\nabla}(\bar{X}_{\kappa}, \theta_{\kappa})  \nonumber\\[2pt]
			&\quad -k_{\tilde{X}}\sum_{i=1}^N \xi_i^{\top}\bar{\psi}(\tilde{X}_i^{-1}\nabla_{\tilde{X}_i} U(\tilde{X}_i, \tilde{\theta}_i))\nonumber \\
			&=-k_X\xi^{\top}\bar{B}(t)\Psi^{\bar{X}}_{\nabla}(x)-k_{\tilde{X}}\xi^{\top}\Psi^{\tilde{X}}_{\nabla}
		\end{align}
		where we made use of the fact $\ad_{\xi_i}\xi_i=\mathbf{0}$. Then, from \eqref{eqn:dot_bar_U}, \eqref{vlofreesumUH_dev} and \eqref{eqn:velocityfreesum_dot_xi_i}, for all $\tilde{x}_c \in \tilde{\mathcal{F}}_c$, the time-derivative of $\tilde{V}$ in \eqref{vlofreeV_def} is given by 
		\begin{align}\label{eqn:dot__widetilde_U}
			\dot{\tilde{V}}(\tilde{x}_c)&=-k_X k_{\theta}\|\Psi_{\nabla}^{\theta}(x)\|^2-k_{\tilde{X}} k_{{\theta}}\|\Psi_{\nabla}^{\tilde{\theta}}\|^2 \nonumber\\ 
			& ~\quad -2k_{\tilde{X}}\Psi^{\tilde{X}\ \top}_{\nabla}(I_N \otimes K_{\hat{X}}) \Psi^{\tilde{X}}_{\nabla}\leq 0.
		\end{align}                            
		This implies that $\tilde{V}(\tilde{x}_c)$ is non-increasing along the flows of \eqref{eqn:vlofreehybsys}. 
		
		Moreover, in view of \eqref{eqn:vlofreehybsys} and \eqref{vlofreeV_def}, for each jump $\tilde{x}_c \in \tilde{\mathcal{J}}_c$, there exists a nonempty set $\mathcal{M}' \subset \mathcal{M}$ or a nonempty set $\mathcal{V}'\subset \mathcal{V}$ such that at least one edge $k\in \mathcal{M}$ or one vertex $i \in \mathcal{V}$ undergoes a jump satisfying $\mu_U(\bar{X}_k, \theta_k) \geq \delta$  or $\mu_U(\tilde{X}_i, \tilde{\theta}_i) \geq \delta$ with $\mu_U$  given by \eqref{mu_U_def},
		one has 
		\begin{align}\label{ineq:velfreV_diff}
			& \tilde{V}(\tilde{x}_c^+)-\tilde{V}(\tilde{x}_c) \nonumber \\
			& \qquad  = k_X\big(\bar{U}(x^+)-\bar{U}(x)\big)\nonumber \\ 
			&\qquad  + k_{\tilde{X}}\sum_{i=1}^N\big(U(\tilde{X}_i^+, \tilde{\theta}_i^+)-U(\tilde{X}_i, \tilde{\theta}_i)\big) \nonumber \\ 
			&\qquad  =-k_X\sum_{k\in \mathcal{M}'}\mu_U(\bar{X}_k, \theta_k)-k_{\tilde{X}}\sum_{i \in \mathcal{V}'}\mu_U(\tilde{X}_i, \tilde{\theta}_i)\nonumber \\ 
			& \qquad  \leq -k_X\sum_{k\in \mathcal{M}'}\delta-k_{\tilde{X}}\sum_{i \in \mathcal{V}'}\delta \leq -\underline{k} \delta
		\end{align}
		where $\underline{k}:=\min\{k_X,k_{\tilde{X}}\}$ and $\delta$ is defined in \eqref{eqn:P_A}. Thus, $\tilde{V}(\tilde{x}_c)$ has strict decrease over the jumps of \eqref{eqn:vlofreehybsys}. Similar to the proof of Theorem \ref{T1}, one can also show that the set $\tilde{\mathcal{A}}$ is stable and every maximal solution of the hybrid closed-loop dynamics \eqref{eqn:vlofreehybsys} is bounded, and the number of jumps is finite.
		
		Next, we show the set $\tilde{\mathcal{A}}$ is globally attractive. 
		Note that every maximal solution to the hybrid system \eqref{eqn:vlofreehybsys} is bounded,  $\tilde{F}(\tilde{x}_c) \in T_{\mathcal{\tilde{F}}_c}(\tilde{x}_c)$ for any $\tilde{x}_c \in \mathcal{\tilde{F}}_c \setminus \mathcal{\tilde{J}}_c$ with $T_{\mathcal{\tilde{F}}_c}(\tilde{x}_c)$ denoting the tangent cone to $\mathcal{\tilde{F}}_c$ at the point $\tilde{x}_c$, and $\tilde{G}(\tilde{x}_c) \subset \mathcal{\tilde{F}}_c \cup \mathcal{\tilde{J}}_c=\mathcal{\tilde{S}}_c$.  Therefore, in terms of \cite[Theorem S3]{goebel2009hybrid}, it is clear that every maximal solution to \eqref{eqn:vlofreehybsys} is complete. Hence, every maximal solution is complete and bounded,
		and thus precompact.
		Using the invariance principle for hybrid systems \citep{goebel2009hybrid}, one has that any solution to the hybrid closed-loop system \eqref{eqn:vlofreehybsys} converges to the largest weakly invariant set contained in $\mathcal{W}: =\{\tilde{x}_c \in \tilde{\mathcal{F}}_c: \Psi_{\nabla}^{\theta}(x)=\mathbf{0}, \Psi_{\nabla}^{\tilde{\theta}}=\mathbf{0}, \Psi^{\tilde{X}}_{\nabla} =\mathbf{0} \}$. For each $i \in \mathcal{V}$, from $\Psi_{\nabla}^{\tilde{\theta}}=\mathbf{0}$ and $\Psi^{\tilde{X}}_{\nabla} =\mathbf{0}$, one has $(\tilde{X}_i, \tilde{\theta}_i) \in \Upsilon_{\tilde{X}}$, with $\Upsilon_{\tilde{X}}:=\{(\tilde{X}, \tilde{\theta}) \in \SE(3) \times \mathbb{R}:
		\nabla_{\tilde{X}}U(\tilde{X}, \tilde{\theta}) = \mathbf{0}, \nabla_{\tilde{\theta}}U(\tilde{X}, \tilde{\theta})=0\}$ denoting the set of all the critical points of $U$. Similar to the proof of Theorem \ref{T1}, one has $(\tilde{X}_i, \tilde{\theta}_i) \in \mathcal{F}_i^o \cap \Upsilon_{\tilde{X}} =\{\mathcal{A}_0\}$.
		Then, from  $\tilde{X}_i \equiv I_4$, it follows that $\dot{\tilde{X}}_i=\mathbf{0}$. This implies  $\xi_i-\beta_i=\mathbf{0}$. From the definition of $\beta_i$ in \eqref{dyncheckX}, it follows from $(\tilde{X}_i, \tilde{\theta}_i) =\mathcal{A}_0$ that $\xi_i=\beta_i=\mathbf{0}$. For each $i\in \mathcal{V}$, from $\xi_i \equiv \mathbf{0}$, one has $\dot{\xi}_i=\mathbf{0}$, which implies that $\dot{\xi}=\mathbf{0}$. 
		Similar to the proof of Theorem \ref{T1}, one can derive that $\Psi_{\nabla}^{\bar{X}}(x)=\mathbf{0}$. 
		Using this fact, together with $\Psi_{\nabla}^{\theta}=\mathbf{0}$ and $\xi=\mathbf{0}$, one can verify that $x_c \in \mathcal{\bar{A}}$. Hence, the largest invariant set of $\mathcal{W}$ is in $\tilde{A}$.
		Finally, one can conclude that the set $\tilde{\mathcal{A}}$ is globally asymptotically stable for the hybrid closed-loop dynamics \eqref{eqn:vlofreehybsys}. This completes the proof.

		\bibliographystyle{apalike}      
		\bibliography{autosam}          
		
	\end{document}